\documentclass[preprint,jgr]{agu2001}
\newcommand{\asap}{Astron. Astrophys.}
\newcommand{\el}{et al.}
\newcommand{\ssr}{Space Sci. Rev.}

\usepackage{epsfig}
\authorrunninghead{LIN ET AL.}

\titlerunninghead{Plasmoids in Reconnecting Current Sheets}

\authoraddr{J. Lin, National Astronomical Observatories of China/Yunnan 
Astronomical Observatory, Chinese Academy of Sciences, Kunming, Yunnan 
650011, China \& Harvard-Smithsonian Center for Astrophysics,
60 Garden Street, Cambridge, MA 02138, USA.
(jlin@ynao.ac.cn)}

\begin{document}

%
%
%
%
%

%
%

\title{Plasmoids in Reconnecting Current Sheets: Solar and Terrestrial 
Contexts Compared}


%
%



\authors{J. Lin, \altaffilmark{1, 2}
S. R. Cranmer, \altaffilmark{2}
C. J. Farrugia \altaffilmark{3}
}

\altaffiltext{1}
{National Astronomical Observatories of China/Yunnan        
Astronomical Observatory, Chinese Academy of Sciences, Kunming, Yunnan
650011, China}

\altaffiltext{2}{Harvard-Smithsonian Center for Astrophysics,
60 Garden Street, Cambridge, MA 02138, USA.}

\altaffiltext{3}{Institute for the Study of Earth, Ocean, and Space,
University of New Hampshire, Durham, NH 03824, USA.}

%
%


\begin{abstract}
Magnetic reconnection plays a crucial role in violent energy conversion 
occurring in the environments of high electrical conductivity, such as the 
solar atmosphere, magnetosphere, and fusion devices. We focus on the 
morphological features of the process in two different environments, the 
solar atmosphere and the geomagnetic tail. In addition to indirect 
evidence that indicates reconnection in progress or having just
taken place, such as auroral manifestations in the magnetosphere and 
the flare loop system in the solar atmosphere,  more direct evidence of 
reconnection in the solar and terrestrial environments is being 
collected. Such evidence includes the reconnection inflow near the 
reconnecting current sheet, and the outflow along the sheet characterized 
by a sequence of plasmoids. Both turbulent and unsteady Petschek-type 
reconnection processes could account for the observations. We also 
discuss other relevant observational consequences of both mechanisms in these
two settings. While on face value, these are two completely different 
physical environments, there emerge many commonalities, for example, an 
Alfv\'{e}n speed of the same order of magnitude, a key parameter 
determining the reconnection rate. This comparative study is meant as a 
contribution to current efforts aimed at isolating similarities in 
processes occurring in very different contexts in the heliosphere, and 
even in the universe.
\end{abstract}

%
%

%

\begin{article}

%
%

\section{Introduction}
Magnetic reconnection is one of the fundamental physical processes that
occur in magnetized plasmas, and manifestations range from laboratory to
magnetospheric plasmas, and to solar and astrophysical plasmas. Reconnection
has as its main characteristic feature the conversion of magnetic field 
energy into other forms of energy, and transfer of plasma and magnetic flux
between separate flux systems: a re-arrangement of magnetic topology. This 
process is associated with phenomena such as heating of the solar corona, 
solar eruptions, coupling between the solar wind and the terrestrial 
magnetosphere, substorms in the geomagnetic tail, and on a more 
speculative level, high-energy phenomena observed in astrophysical systems
[e.g., see {\it Priest and Forbes} 2000].

Because these phenomena occur in environments of high electrical 
conductivity, the energy conversion process is usually confined to a 
small, local region, such as the X-type neutral point, the current sheet, 
the quasi-separatrix layer, and so on. Among these types of magnetic 
neutral regions, the X-point is probably the most basic configuration from 
which the others could develop. For example, a current sheet could
develop either through the collapse of an X-point [{\it Dungey} 1953] or 
through the severe stretching of the magnetic field around an X-point [{\it 
Forbes and Isenberg} 1991]; and a quasi-separatrix layer [{\it 
D\'{e}moulin et al.} 1996] might be produced by skewing the magnetic 
fields on both sides of an X-point.

At the  current sheet separating the shocked solar wind and the terrestrial 
magnetosphere (the magnetopause), magnetic reconnection can transfer 
solar wind momentum and energy into the magnetosphere on the dayside. Figure 
\ref{fig:magnetosphere} displays a schematic of the topological changes
that are initiated by magnetic reconnection between solar wind 
field and terrestrial field on the dayside (stage 1). Reconnected (or 
``open") field lines are then dragged by the shocked solar wind flow 
into the nightside magnetosphere, where they form the tail lobes (stage 
2). Addition of magnetic flux from the dayside to the nightside increases 
the energy stored in the tail. A thin current sheet forms in the tail and at 
substorm onset reconnection takes places there (stage 3), heating plasma 
and jetting it in bursty fashion both earthward [{\it Baumjohann et al.}, 
1990; {\it Angelopoulos et al.}, 1992] (stage 4) and tailward (stage 5) 
at high speeds. On the earthward side, reconnected field lines rapidly 
snap back toward the Earth due to magnetic tension, constituting a 
process known as the shrinkage of field lines [e.g., see {\it \v{S}vestka 
et al.} 1987; {\it Lin et al.} 1995a; {\it Forbes and Acton} 1996; {\it 
Lin} 2004; {\it Reeves et al.} 2008]. These field lines eventually return 
to the dayside and replace the eroded magnetic flux.

Magnetic reconnection and a general convective cycle, including 
frontside and nightside reconnections, in the process described above were 
proposed for the first time by {\it Dungey} [1961]. Later, the substorm 
model was developed, particularly by {\it McPherron et al.} [1973], {\it 
Hones} [1977], and {\it McPherron} [1991]. (See also {\it Baker et al.} 
[1996] for a more recent review.) As emerges from the discussions above, 
the substorm is an explosive energy release and is accompanied by a 
reconfiguration of the near-tail magnetic field (dipolarization) and 
other phenomena to be discussed later. If the magnetosphere is immersed 
in a southward-pointing interplanetary magnetic field of long (many 
hours) duration, such as what typically happens during passage of 
interplanetary coronal mass ejections (ICMEs), then repetitive substorms 
[{\it Farrugia et al.} 1993a] occur at intervals of about 2--3 hours [{\it 
Huang et al.} 2003].

The formation and development of a current sheet and/or a 
quasi-separatrix in the solar magnetic field is usually associated with 
solar eruptive processes, including solar flares, eruptive prominences, 
and coronal mass ejections (CMEs). In fact, substorms have been called the
flares in the magnetosphere; i.e., loosely speaking, due to the 
similarity and explosive character in the energy conversion processes, 
substorms may be considered as the magnetospheric counterparts of
solar eruptions. As we show below, a comparison of reconnection in the solar 
and terrestrial contexts reveals many common features.

Table \ref{tbl:VA} lists some important parameters for the solar 
atmosphere and for the magnetosphere. Those for the solar atmosphere can 
be found in the book by {\it Priest} [1982], those for the low latitude 
dayside magnetosphere were taken from the {\it in situ} measurements 
reported by {\it Paschmann et al.} [1986], those for the middle distant 
tail lobe and the plasma sheet are from {\it Slavin et al.} [1985], and 
those for the near Earth sheet are from {\it Baumjohann et al.} [1989]. 
We notice the same orders of magnitude of large $V_{A}$ and small $\beta$ 
in the solar corona and in the tail lobes of the magnetosphere, but 
small $V_{A}$ and larger $\beta$ in the solar photosphere, in the dayside 
magnetopause, and in the central plasma sheet in the geomagnetic tail. 
Those in the central sheet (e.g., see {\it Baumjohann et al.} [1989]) 
suggest a rapid increase of the reconnection rate when low density, high 
Alfv\'{e}n speed, lobe field becomes reconnected (e.g., {\it Hesse et 
al.} [1996]). However, the corresponding values for the solar counterpart 
of the geomagnetic tail sheet are not available due to the lack of {\it 
in situ} measurements in the solar environment. On the other hand, the 
low $V_{A}$ and high $\beta$ in the solar photosphere and in the dayside
magnetopause have different consequences that will be discussed shortly.

So even though Table \ref{tbl:VA} indicates some similarities of the 
current sheet in the magnetosphere to that in the solar corona, the levels 
of our understanding of the details of magnetic reconnection in the 
magnetosphere and in the solar corona are very disparate. In the 
magnetosphere, extensive {\it in situ} measurements reveal tremendously 
rich information of the internal features of the current sheets and the 
magnetic reconnection processes at the magnetopause (e.g., {\it Rijnbeek 
et al.} [1989]; {\it Sonnerup et al.} [1981]; {\it Phan et al.} [2000]; 
{\it Paschmann} [2005]), and in the magnetotail [e.g., {\it Murata et al.} 
1995; {\it Baker et al.} 1985, {\it Nakamura et al.} 2006]. In the solar 
corona, on the other hand, directly observing the current sheet turns out 
to be very hard, if not impossible. Difficulties stem from two aspects: 
the low level of emission from the sheet plasma compared to that of other 
nearby bright features (see the introduction of {\it Ko et al.} [2003] for 
a brief review), and the possible thinness of the sheet.

In theory, it has often been stated that the sheet is too thin to be
observable since its thickness is believed to be limited by the Larmor
radius of particles that is only a few tens of meters in the coronal
environment [{\it Litvinenko} 1996; {\it Wood and Neukirch} 2005; and 
references therein]. Another reason that might have discouraged direct 
observations of the current sheet could be ascribed to the ``circuit 
model" of eruptions developed by {\it Martens and Kuin} [1989], in which the 
theoretical current sheet length is short. Not until the theoretical 
works of {\it Forbes and Lin} [2000], {\it Lin and Forbes} [2000], and
{\it Lin} [2002], as well as a series subsequent observational ones 
[{\it Ciaravella et al.} 2002; {\it Ko et al.} 2003; {\it Raymond et al.} 
2003; {\it Sui and Holman} 2004; {\it Sui et al.} 2004; {\it Webb et al.} 
2003; {\it Lin et al.} 2005; {\it Bemporad et al.} 2006; {\it Lin et al.} 
2007], have we started realizing that a fairly long current sheet could 
develop in the major eruptive process.

In this work we undertake a comparison of the reconnection processes in 
these two environments, putting emphasis on geomagnetic substorms, on the
one hand, and solar flares, on the other. Such a comparison consists of two 
aspects. First, we look into the detailed internal structures of the 
reconnecting sheets observed in the geomagnetic environment and in the 
solar corona; second, we compare two theoretical mechanisms that may 
account for the observed features of the reconnecting current sheet. This 
brings the known results for magnetic reconnection obtained from the 
investigations of various communities together, and helps people of 
different research areas to better understand the work and the progress made 
by others.

%
%


%
%

\section{Reconnection-related plasmoids (plasma blobs) 
in the solar and terrestrial environments}
In this part of work, we go through a set of observational results 
regarding plasmoids (plasma blobs) observed in different environments. 
The term ``plasmoid" is usually used to describe flowing features in the 
geomagnetic tail current sheet [e.g., see {\it Murata et al.} 1995 and 
references therein]. The term ``plasma blob", on the other hand, is used 
for the similar features observed in the CME/flare sheets [see {\it Ko et 
al.} 2003]. Observed morphological features and the possible causative 
mechanisms suggest that plasma blobs and plasmoids are likely the same 
object, produced by reconnection.

\subsection{Reconnection Processes Observed in the \\ Magnetotail}
\begin{figure}
\noindent\includegraphics[width=26pc]{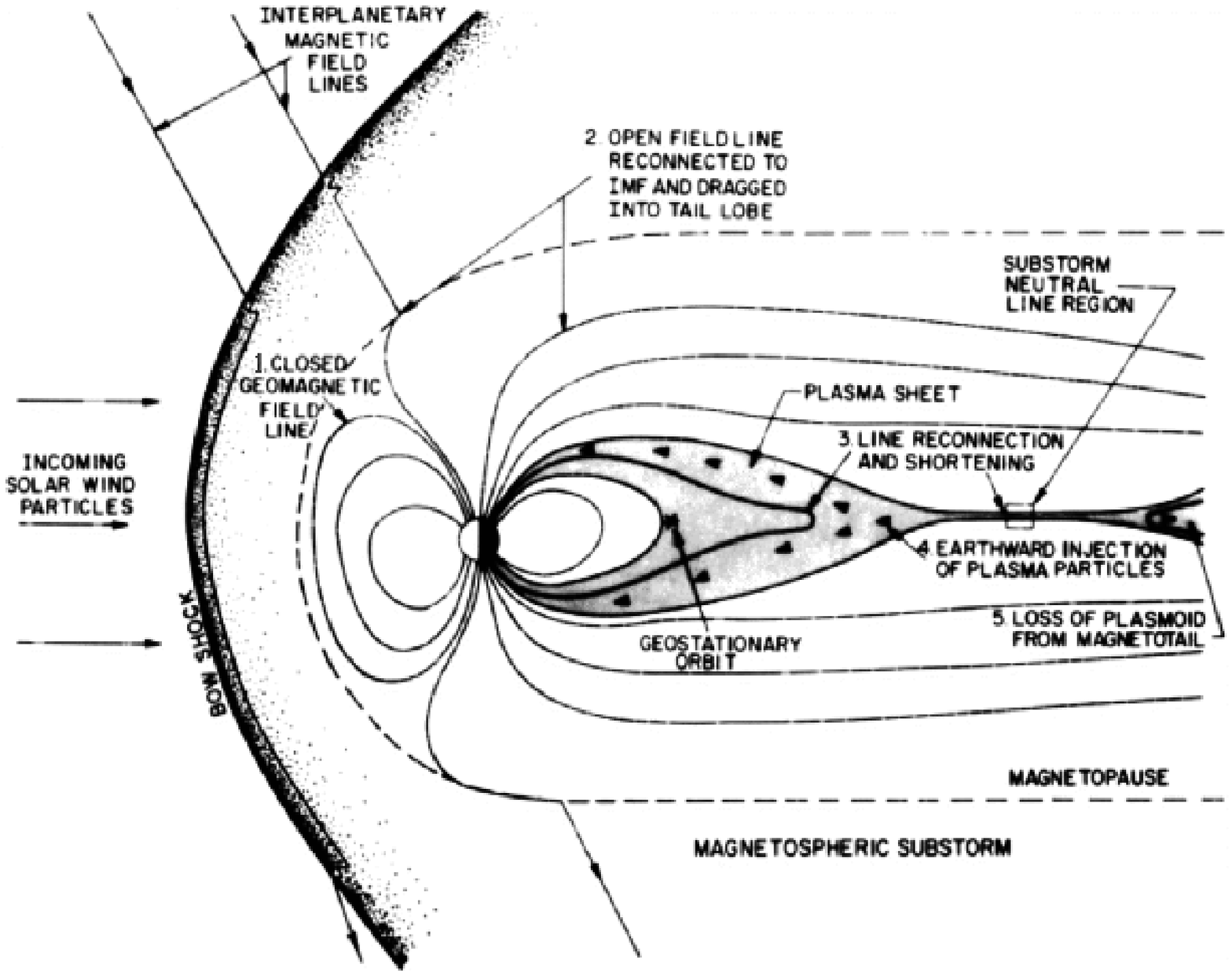}
\caption{Schematic diagram of magnetic reconnection processes occurring 
in the magnetosphere. Interplanetary magnetic field in the solar 
wind reconnect with the terrestrial field near the front of the 
magnetosphere (stage 1) and are dragged back to form the tail lobes, sending 
plasma and magnetic energy there and forming the neutral plasma 
(current) sheet (stage 2). When the sheet becomes thin enough as a result 
of the increase in the magnetic pressure in the lobes, magnetic 
reconnection commences and releases the magnetic energy to heat and 
accelerate magnetotail plasma (stage 3), sending plasmoids towards (stage 
4) and away from the Earth (Stage 5). From Figure 1 of {\it Baker et al.} 
[1987].}
\label{fig:magnetosphere}
\end{figure}

The existence of the Earth's geomagnetic tail and its embedded current 
sheet was confirmed by extensive {\it in situ} measurements of the 
magnetic field of the Earth at distances greater than 7 Earth radii in the 
1960s [e.g, see {\it Ness} 1965], and has been observed hundreds of Earth 
radii from the Earth by spacecraft. The discovery of the neutral sheet in 
the magnetotail is an important aspect in determining the interaction of 
the super Alfv\'{e}nic solar wind with the Earth's magnetic field. The 
dynamics of this sheet may thus play an essential role in geomagnetic 
phenomena [{\it Ness} 1965]. The geomagnetic tail consists of magnetized 
plasma in the tail lobes overlaying the plasma sheet, in which the 
current sheet is embedded separating two oppositely-directed magnetic 
fields. The geomagnetic tail results from the flow of the solar wind 
around the magnetosphere. Reconnection of the magnetospheric and the 
interplanetary magnetic fields starts on the dayside. Open flux is 
transferred to the nightside where it is stored during the substorm 
growth phase until released again to the dayside at the substorm onset (see 
also Figure \ref{fig:magnetosphere} for the cycle).

Dayside reconnection may be quasi-steady [e.g., {\it Sonnerup et al.} 
1981] or time-dependent. The latter form was discovered by {\it Russell 
and Elphic} [1978] at about the same time as observational evidence was 
being shown for steady reconnection. Time dependent reconnection features 
on the dayside are known as flux transfer events (FTEs). These are tied to 
various signatures, chief among these is a bipolar variation lasting 
1-2 min of the magnetic field component normal to the magnetopause (see 
reviews by {\it Elphic} [1995] and {\it Farrugia et al.} [1988]). The low 
$V_{A}$ and high $\beta$ in this region (see Table \ref{tbl:VA}) imposes 
restrictions on the rate of magnetic reconnection and thus the energy 
conversion. So particle acceleration and plasma heating at the dayside 
magnetopause are not as obvious as in the tail region {\it Cowley} [1980]. 
Instead reconnection here primarily leads to the storage of magnetic 
reconnection converted from the solar wind kinetic energy (see Figure 
\ref{fig:magnetosphere}).

As indicated by Figure \ref{fig:magnetosphere}, after reconnection on the 
dayside open field lines are dragged to the nightside forming the tail 
lobes. This results in a considerable increase in the energy stored in the 
tail. To balance the dayside reconnection during relatively quiet times, 
nightside reconnection is needed, which is assumed to take place in the 
far tail forming the quiet time plasma sheet. During the rapid accretion 
of magnetic energy in the tail lobe during disturbed conditions, 
however,  the plasma sheet closer to Earth (at $10-20$ $Re$) becomes 
thinner.  When it becomes sufficiently thin, a near-Earth neutral line is 
created. At the neutral line plasma is heated and jets in both earthward 
and tailward directions. In earthward direction, closer to the Earth than 
the neutral sheet the field lines become more dipolar through magnetic 
tension, injecting energetic particles near local midnight in the process. 
Strong auroral effects are produced at the magnetic footprints of these 
field lines (e.g. {\it Sergeev et al.} [1999]; {\it Sandholt et al.} 
[2001]); and in the tailward direction, a ``magnetic island" or plasmoid 
is propelled at high speeds.

The term ``plasmoid" was first introduced by {\it Hones et al.} [1977] who 
took it from the laboratory literature where the pulse of plasma emitted 
into a chamber to conduct experiments was termed a plasmoid. Magnetotail 
plasmoids as regions of typically reduced magnetic field strength (except 
when they contain a strong core field), surrounded by a traveling 
compression region, were discovered in the ISEE data in the mid-1980s 
[{\it Hones et al.} 1984; {\it Slavin et al.} 1984]. They are large (about 
tens of $R_{E}$) closed magnetic loops which propagate rapidly downtail 
during a substorm event [{\it Hones et al.} 1984]. {\it Slavin et al.} 
[1984] described traveling convection regions observed by ISEE in the 
distant tail consequent upon substorm activity nearer Earth and identified 
them as lobe signatures of plasmoids traveling at a few 10$^{2}$ km 
s$^{-1}$ downtail. The release of plasmoids in the near-Earth tail at 
substorm expansion phase was inferred by {\it Hones} [1979].

Features of the distant neutral line (DNL) are also summarized in Figure 
\ref{fig:magnetosphere}. The DNL was first demonstrated with actual 
measurements from ISEE 3 by {Slavin et al.} [1985]. It was then examined 
in a series of further studies using Geotail data by {\it Nishida et al.} 
[1994; 1998; and references therein]. Reconnection at the DNL appears 
relatively continuous and moderate in terms of its outflow speed, perhaps 
due to the fact that the lobes are filled with diffuse plasma by $x$ (GSM) 
$\sim 100 Re$ from the ``mantle" that lowers the Alfv\'{e}n speed.

Magnetic island structures in the magnetotail have been frequently 
detected by {\it in situ} measurements either by direct encounter or 
by remote sensing the perturbations they excite in the surrounding field 
and plasma [{\it Slavin et al.} 1984] during geomagnetic substorms, and 
the flow speed could be as high as several hundreds of km/s [{\it Ho and 
Tsurutani} 1997]. Reports in a series of recent works on the multiple 
X-line reconnection (e.g., {\it Slavin et al.} [2003]; {\it Slavin et 
al.} [2005]; and {\it Eastwood et al.} [2005]) based on the observations 
made by the four Cluster spacecrafts confirmed the magnetic island 
structure inside the current sheet. These observations indicated the 
occurrence of the multiple X-line reconnection in the magnetotail proposed 
by {\it Lee and Fu} [1985] and {\it Fu and Lee} [1985].

These authors found plasmoid features moving both earthward and tailward, 
and argued that they are the products of the multiple X-line 
reconnection. The rate of reconnection at each X-line is not necessarily 
the same. Instead, one of these X-lines becomes dominant where the fastest 
reconnection takes place. {\it Eastwood et al.} [2005] pointed out that 
although numerical and experimental evidence had been observed to suggest 
the filamentation of the current sheet and multiple X-line reconnection, 
the data from Cluster and related investigations presented the 
observational evidence of the relevant features in the magnetotail for the 
first time. On the other hand, what determines the location of the 
dominant X-line is unknown. It probably depends on the adjacent magnetic 
field and the plasma environment.

Some other interesting characteristics of plasmoids are that plasmoids 
observed to move tailward are more than those moving earthward [{\it 
Moldwin and Hughes} 1991], and that the tailward motion is faster than 
the earthward motion [{\it Slavin et al.} 2003]. This is probably due to 
the fact that the earthward flow is deflected and stopped near the cusp 
point by the closed magnetic field lines near the Earth, as described by 
{\it Reeves et al.} [2008] in a recent work related to the present one. 
Because of mathematical complexity, the behavior of the plasmoid 
approaching the cusp region was tentatively studied via the analytic 
approach in only two works so far [i.e., {\it Semenov and Lebedeva} 1991; 
{\it Lin et al.} 1995b], and no numerical experiment focusing on the 
behavior of the plasmoid near that area has yet been reported.

Another important consequence of reconnection is the generation of fast   
plasma flows, which for a long time have been used as the major indicator
of the occurrence of reconnection and for the identification of the 
location of the reconnection site. On the dayside, accelerated plasma flows 
have been extensively used to infer the presence of reconnection [see {\it 
Paschmann et al.} 1979; {\it Sonnerup et al.} 1981] in view of the 
difficulty in showing a non-zero normal component. In the tail, bulk 
plasma acceleration associated with the formation of the slow mode shock 
that has been identified (e.g., see {\it Feldman et al.} [1984]; {\it 
{\O}ieroset et al.} [2000];  and {\it Sonnerup et al.} [1987]) implicitly 
suggests that the Petschek-type reconnection is in operation. Large-scale 
hybrid simulations of the magnetotail undergoing reconnection were 
conducted, and significance of the slow mode shocks was confirmed [{\it 
Krauss-Varban and Omidi} 1995]. But in the quasi-steady state framework, 
not all the observed and simulated features, such as the slow mode shock, 
gave a coherent and consistent picture [{\it Feldman et al.} 1985; {\it 
Krauss-Varban and Omidi} 1995]. From this time-dependent properties of 
reconnection process in reality may be inferred [e.g., {\it Pudovkin and 
Semenov} 1985].

\subsection{Reconnection Processes Observed in Solar Eruptions}
\begin{figure}
\noindent\includegraphics[width=30pc]{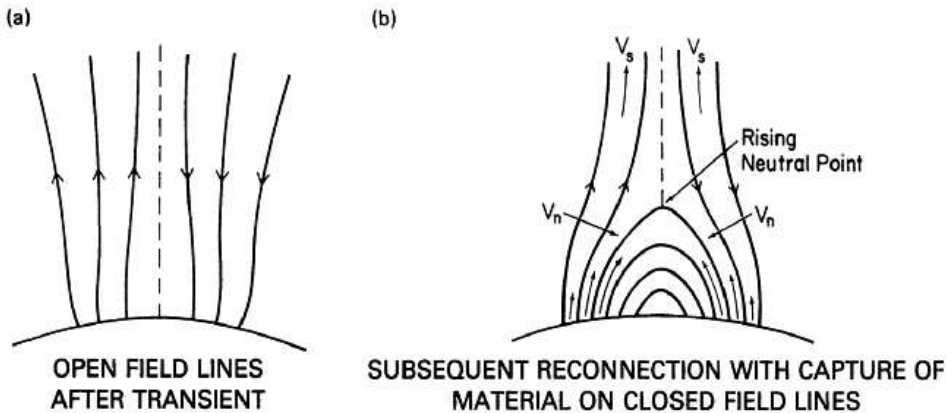}
\caption{The two-ribbon flare model by {\it Kopp and Pneuman} [1976]. (a) 
The magnetic field is pushed open by an eruption and a current sheet 
separates two anti-parallel magnetic field lines. (b) The opened 
configuration relaxes into a closed, nearly potential  field via magnetic 
reconnection in the current sheet. This process produces two bright and 
separating flare ribbons on the solar disk, and a continually growing 
flare loop system in the corona. From {\it Kopp and Pneuman} [1976].} 
\label{fig:Kopp-Pneuman}
\end{figure}

\begin{figure}
\noindent\includegraphics[width=20pc]{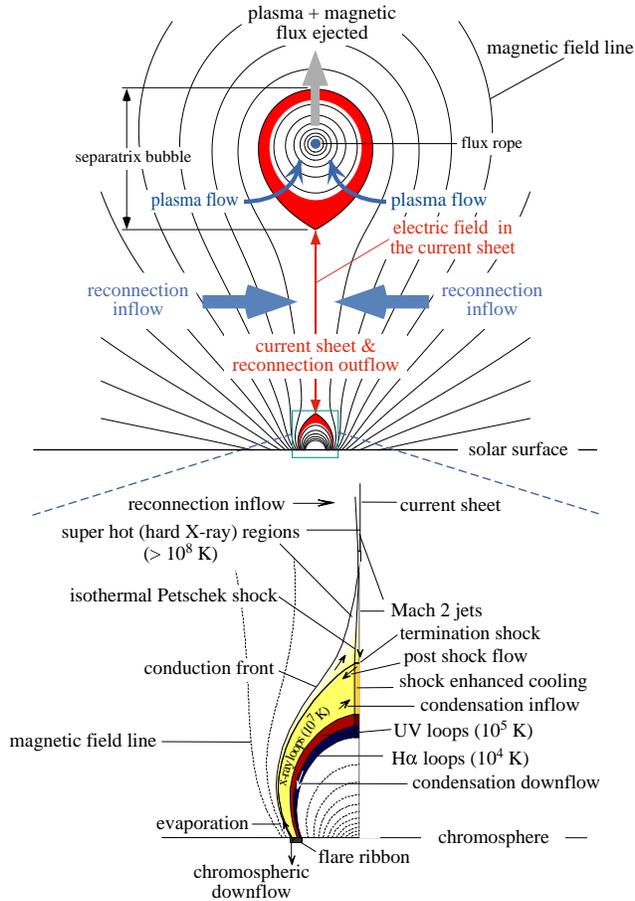}
\caption{Schematic diagram of a disrupted magnetic field that forms in an
eruptive process. Colors are used to roughly denote the plasma layers 
in different temperatures. This diagram incorporates the two-ribbon
flare configuration of {\it Forbes and Acton} [1996] and the CME 
configuration of {\it Lin and Forbes} [2000], and provides a 
comprehensive description of how various manifestations in the solar 
eruption are related to one another. The morphological features of the 
disrupting magnetic field in the event studied by {\it Lin et al.} [2005] 
nearly duplicated those of this diagram (cf. Figure 5 of {\it Lin et 
al.} [2005]).}
\label{fig:CME-flare}
\end{figure}

There are two solar counterparts of the geomagnetic tail: One is the 
current sheet that extends from the top of flare loops and connects to the 
associated CME (e.g., see {\it Lin et al.} [2004]); and another one is 
embedded in the ``helmet streamers". Similarities  of these two objects 
in morphology can be seen from both observations (e.g., see {\it Priest} 
[1982]) and numerical experiments (e.g., see {\it Birn et al.} [2003]). The 
current sheet configuration related to flares and CMEs cannot exist for 
long, and is created presumably by the disrupting magnetic field as a 
result of the loss of equilibrium in the system (e.g., see {\it Forbes 
and Isenberg} [1991]; {\it Forbes} [2000 and 2007]; and references 
therein). Such a scenario was suggested for the first time by {\it 
Carmichael} [1964], and was later developed into the well-known 
Kopp-Pneuman model [{\it Kopp and Pneuman} 1976], or more generally, the 
CSHKP model [e.g., see {\it \v{S}vestka and Cliver} 1992] (see Figure 
\ref{fig:Kopp-Pneuman}). It was further developed by {\it Martens and 
Kuin} [1989] and {\it Lin and Forbes} [2000] to the catastrophe model of 
the solar eruption in which the eruptive prominence evolves into the CME 
connecting to the associated flare below by a long current sheet (see 
Figure \ref{fig:CME-flare}, and also {\it Lin et al.} [2003] for other 
models).

In this model, continuous mass motions and magnetic reconnection in the 
photosphere successively displace the footpoints of the coronal magnetic 
field, changes the shape of the field, converts the kinetic energy of 
mass motions into the magnetic energy and transports it into the corona. 
The low $V_{A}$ and high $\beta$ in the photosphere limits the rate of 
magnetic reconnection there such that no apparent heating and 
acceleration of plasma takes place (refer to Table \ref{tbl:VA}). 
Instead the magnetic reconnection in the photosphere plays the role in 
transporting magnetic flux only. Here we note, again, the process of 
converting the kinetic energy from the low $V_{A}$ and high $\beta$ 
region (the photosphere) into magnetic energy stored in the high 
$V_{A}$ and low $\beta$ region (the corona). Eventually, when the 
stored energy in the corona exceeds the threshold, the magnetic field 
loses the equilibrium and disrupts. Magnetic field lines are severely 
stretched to form an effectively open magnetic configuration including a 
neutral sheet separating magnetic fields of opposite polarity. Magnetic 
reconnection occurring inside the current sheet creates the growing flare 
loops in the corona and the separating flare loops on the solar disk 
(e.g., see {\it Forbes} [2000]; {\it Klimchuk} [2001]; {\it Priest and 
Forbes} [2002]; {\it Lin et al.} [2003]; {\it Birn et al.} [2003 and 2006]; 
{\it Forbes} [2007], and Figure \ref{fig:CME-flare} as well). When magnetic 
reconnection occurs in this current sheet, the high $V_{A}$ and low 
$\beta$ in the corona determines that the stored magnetic energy is 
released in a violent fashion.

Because the current sheet associated with flare and CME can only form and 
exist in the eruptive process, continuous reconnection analogous to that 
occurring at DNL in the Earth's tail ceases as the eruption ends. A 
typical eruptive process usually lasts tens of hours (e.g., see {\it 
\v{S}vestka and Cliver} [1992]), which may also be the lifetime of the 
corresponding continuous reconnection. The current sheet in the helmet 
streamer, on the other hand, can last much longer. The helmet 
streamer is the huge, long-lived, radially oriented structure that 
extends from the base of the corona out to several solar radii (e.g., see 
{\it Aschwanden} [2005], p. 9). The lower part of it contains closed 
magnetic field spanning over a magnetic neutral line on the solar 
surface, while the upper part turns into an open magnetic field joining 
the closed one in the region of the cusp-geometry from which the gas 
pressure starts to exceed the magnetic pressure outward. The long-lived 
property of the helmet streamer allows magnetic reconnection in the 
sheet to continue for a while analogous to that in the Earth's tail.

The helmet streamers will be discussed in the next section, and in 
this section we focus here on the CME/flare current sheet. Recent research 
indicates that the closed magnetic field does not necessarily become 
fully open as the Kopp-Pneuman-type eruption occurs. Instead, the magnetic 
structure is severely stretched and a current sheet forms [e.g., {\it Lin 
and Forbes} 2000; {\it Forbes and Lin} 2000]. With dissipation in the 
current sheet, the stretched magnetic field starts to reconnect, 
producing new closed field lines both below and above the current sheet 
[{\it Lin et al.} 2004], so the pre-existing closed magnetic field never 
becomes fully open (see Figure \ref{fig:CME-flare}).

Related to the above theoretical results, a series of observations 
[{\it Ciaravella et al.} 2002; {\it Ko et al.} 2003; {\it Raymond et al.} 
2003; {\it Sui and Holman} 2004; {\it Sui et al.} 2004; {\it Webb et al.} 
2003; {\it Lin et al.} 2005; {\it Bemporad et al.} 2006] provided evidence 
for the formation and development of the long current sheets in major 
eruptions. Those of {\it Ko et al.} [2003], {\it Lin et al.} [2005], {\it 
Bemporad et al.} [2006] , and {\it Lin et al.} [2007] further specified 
and analyzed the internal features of the long current sheets connecting 
solar flares to the associated CMEs. The most significant features are 
plasma blobs flowing away from the Sun along the sheets. Similar features 
flowing toward the Sun on the top of flare loops had been observed and 
identified with the reconnection outflow earlier [{\it \v{S}vestka et al.} 
1998; {\it McKenzie and Hudson} 1999; {\it Sheeley and Wang} 2002; {\it 
Asai et al.} 2004; {\it Sheeley et al.} 2004], and numerical experiments 
for solar eruptions showed the blobs and flows moving in both directions 
[{\it Y. Fan} 2005, private communications; {\it Riley et al.} 2007]. 

The term ``plasma blob" is not as well-known as plasmoid in either the 
solar physics or the geophysics communities since it was never used 
before {\it Ko et al.} [2003] who applied it to describing the flowing 
features along the reconnecting current sheet in an event producing a 
fast CME and an X-class flare (see their Figures 7 and 18). Subsequently, 
{\it Lin et al.} [2005] and {\it Riley et al.} [2007] observed
similar features both in observations of another major event and in
numerical experiments (see Figure 3 of {\it Lin et al.} [2005] and Figure 3 
of {\it Riley et al.} [2007]). Following {\it Ko et al.} [2003], they 
also utilized ``plasma blobs" to describe these features. Comparison of 
the behavior and properties of plasma blobs observed in the solar eruption 
with those of plasmoids detected in the geomagnetic tail (see Figure 3 of 
{\it Hones} [1977], Figure 1.6 of {\it Birn and Priest} [2007], and Figure 
\ref{fig:magnetosphere} of this work) suggests the identity of two objects 
appearing in different environments.

We also note that the term ``flux rope" often appears in the literature of 
geophysics as well (e.g. {\it Slavin et al.} [1995]). The relevant 
structure is very similar to that of plasmoid flowing inside the 
geomagnetic tail, but {\it Sibeck et al.} [1984] also noticed several 
differences between the two. One of these important differences is the 
presence of an axial magnetic field inside the geo-flux-rope and the 
magnetic connection of the rope to the Earth (e.g., see discussions of 
{\it Moldwin and Hughes} [1991], and a recent numerical experiment by 
{\it Birn et al.} [2004]). This feature is quite like that of the 
flux rope in the solar environment, which includes a core field and is 
anchored to the Sun at its both ends (e.g., see {\it Lin et al.} [2004]).

{\it Lin et al.} [2007] and {\it Riley et al.} [2007] identified plasma 
blobs in the current sheet with the magnetic islands resulting from 
the tearing mode instability (turbulence). But such blob-like or 
island-like structures inside the reconnecting current sheets may also be 
ascribed to an alternative version of the energy conversion: the 
time-dependent Petschek-type reconnection [e.g., {\it Pudovkin and 
Semenov} 1985; {\it Priest and Forbes} 2000]. The theory of such a 
process was developed from the steady-state analysis of reconnection 
performed by {\it Petschek} [1964], and was then used to explain FTEs 
observed at the dayside magnetopause [see also {\it Rijnbeek et al.} 
1989; {\it Semenov et al.} 1992; and references therein]. The slow mode 
shock has not yet been seen in the laboratory due to experimental 
difficulties [e.g., see {\it Hada and Kennel} 1985]. A possible approach 
to relating the observed plasma blobs to the reconnection outflow regions 
surrounded by the slow mode shock was recently discussed by {\it Lin et 
al.} [2007].

Compared with those detected in the magnetotail, the solar counterpart of 
the reconnection flows and the related features were not observed until 
{\it \v{S}vestka et al.} [1998] noticed a fan of spikelike ray structures 
above a group of flare loops observed by {\it Yohkoh}. Then, the apparent 
mass motion toward the Sun above the flare loop system in the long 
duration event of 1999 January 20 was reported by {\it McKenzie and 
Hudson} [1999]. Comparison with the standard model of two-ribbon flares 
[e.g., see {\it Forbes and Acton} 1996] suggests fitting of the 
observation to the theory. {\it McKenzie and Hudson} [1999] found that the 
late-phase downward motion was in the form of soft X-ray dark voids having 
speeds that vary from 100 km s$^{-1}$ to 200 km s$^{-1}$, the 
temperature in this region reached up to $9.1\times 10^{6}$ K, and the 
density was a few times $10^{9}$ cm$^{-3}$. These data indicate that the 
dark X-ray voids are the blobs, or magnetic islands in the reconnecting 
current sheet. Another 11 examples that showed similar Sun-ward flows 
during the long-duration events observed by {\it Yohkoh} SXT were also 
reported by {\it McKenzie} [2000]. The speeds of those flows ranged from 
50 km s$^{-1}$ to 500 km s$^{-1}$, and all the events were associated with 
CMEs.

\begin{figure}
\noindent\includegraphics[width=20pc]{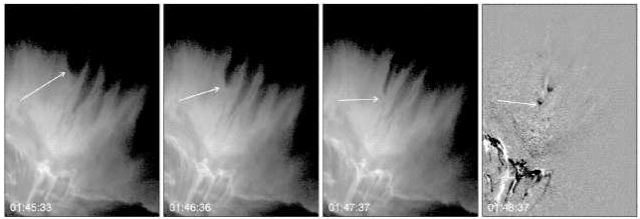}
\caption{{\it TRACE} 195 {\AA} filtergrams of the 2002 April 21 west-limb 
flare showing Fe {\small XII} postflare loops (lower left) and the Fe 
{\small XXIV} plasma cloud (center) penetrated by dark tadpole-like 
inflows (upper right). The fourth panel is a difference image showing the 
change between the images at 0147 and 0148 UT. These images have been 
rotated so that the solar limb is approximately horizontal. The arrow 
refers to a ``tadpole" flowing toward to the Sun. The vertical dimension 
of each panel is approximately $1.17\times 10^{5}$ km. From {\it Sheeley et 
al.} [2004].} \label{fig:tadpole}
\end{figure}

Subsequently, a series of plasma downflows above the post-flare loops
were successively observed. {\it Sheeley and Wang} [2002] reported the 
coronal downflows observed with LASCO 2$-$6 $R_{\odot}$ from the 
heliocenter. The maximum velocities of individual downflows varied from 
50 km s$^{-1}$ to 100 km s$^{-1}$. The well-known TRACE event that 
occurred on April 21, 2002 gave rise to an X1.2 class flare and a very 
fast CME ($\ge 2500$ km s$^{-1}$). It was also well observed by the 
instruments on SOHO, such as LASCO, SUMER, EIT, and UVCS [{\it Innes {\el}} 
2003; {\it Raymond {\el}} 2003; {\it Sheeley {\el}} 2004]. The downward flow 
features were observed at the early stage of the eruption, and both the 
speed and temperature of the plasma flow were fairly high (up to 1000 km 
s$^{-1}$ and a few times 10$^{7}$ K, respectively). {\it Sheeley {\el}} 
[2004] employed a technique developed to study motions in the outer 
corona [see {\it Sheeley {\el}} 1999] to track the plasma downflows 
displayed by this event (Figure \ref{fig:tadpole}). What is nice 
about their technique is that it helps manifest in a very clear way 
several important features related to magnetic reconnection. These 
features include the fast reconnection outflow ($100 - 600$ km s$^{-1}$) 
towards the flare loops and the significant deceleration ($\sim$ 1500 m 
s$^{-2}$) of the flows at the top of the flare loop system, which is 
highly suggestive of the reconnection outflow encountering the closed 
field line region in the magnetotail context (refer to those nightside 
closed field lines in Figure \ref{fig:magnetosphere} and those below the 
current sheet in Figure \ref{fig:Kopp-Pneuman}$b$).

{\it Asai {\el}} [2004] presented a detailed study of downward motions above
flare loops observed in the July 23, 2002 event. This event produced an 
X4.8 flare and an energetic CME ($\sim$ 2600 km s$^{-1}$), and was well
observed by TRACE, RHESSI, NoRH, UVCS, and LASCO [{\it Emslie {\el}} 2003;   
{\it Raymond {\el}} 2003]. During this event, the downflows above the 
post-flare loops were seen not only in the decay phase but also in the 
impulsive and main phases, and showed clear correlation to the 
non-thermal emissions in microwaves and HXRs. Magnetic reconnection was 
thought to account for these characteristics. Most recently, a limb flare 
was observed by SOHO/SUMER, TRACE, and RHESSI, and the sunward reconnection 
outflows consisting of hot plasma along the flare loops was reported by 
{\it Wang et al.} [2007].

\begin{figure}
\noindent\includegraphics[width=25pc]{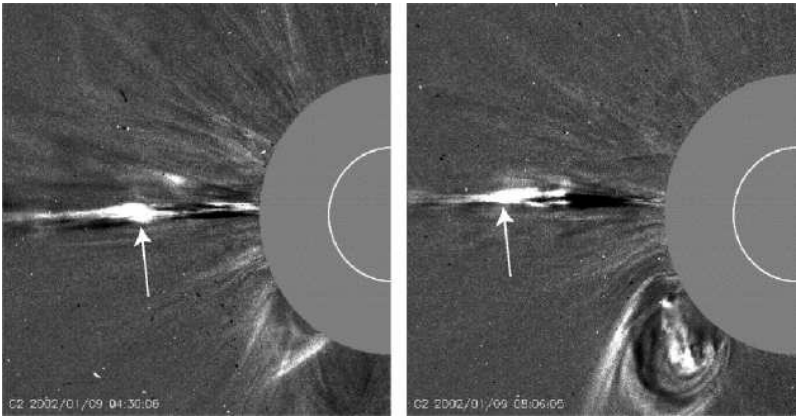}
\caption{Running difference images of two examples of the plasma blobs 
observed by LASCO C2 flowing away from the Sun along the current sheet. 
From {\it Ko et al.} [2003].} 
\label{fig:blob1}
\end{figure}

\begin{figure}
\noindent\includegraphics[width=25pc]{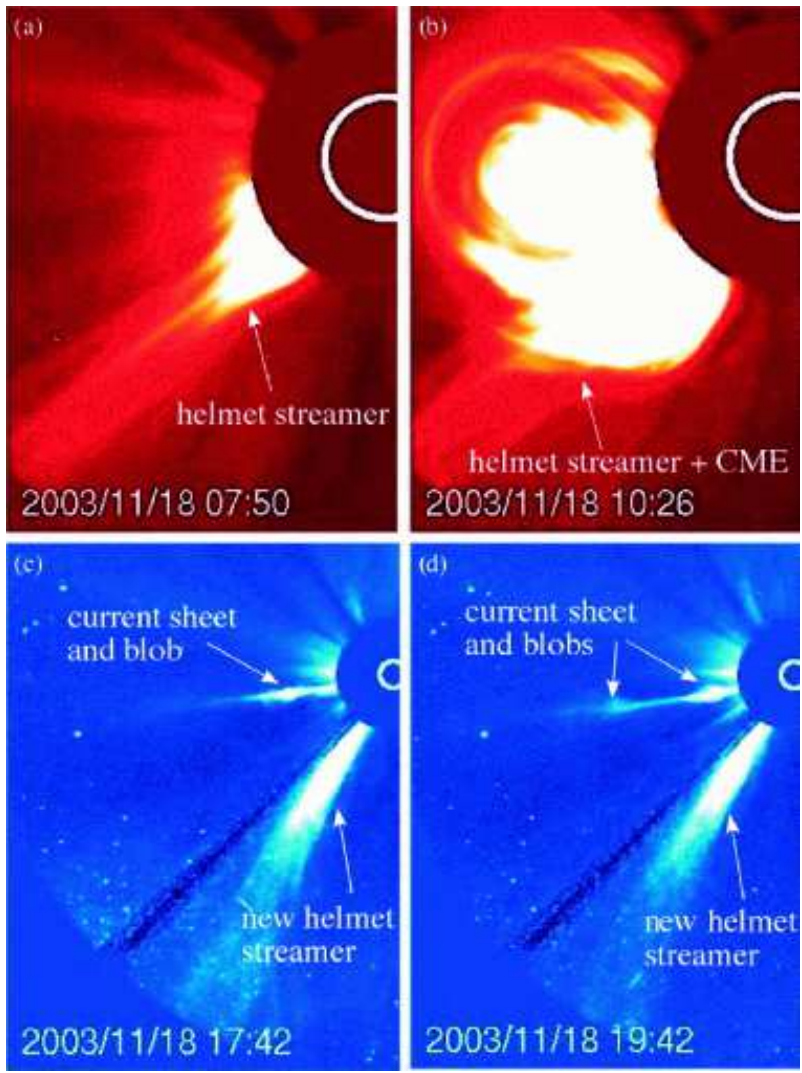}
\caption{A fast CME was observed by LASCO C2 (panels a and b), and a long 
thin current sheet left behind by the CME can be seen in LASCO C3 images 
with two well recognized plasma blobs flowing away from the Sun (panels c 
and d). From {\it Lin et al.} [2005].}
\label{fig:blob2}
\end{figure}

Because of the relatively small scale and the low emission measure of the
current sheets in the solar eruption, the magnetic reconnection outflow    
moving away from the Sun was not noticed until a major eruptive process 
was observed by {\it Ko et al.} [2003] that developed a typical 
CME$-$current sheet$-$flare loop system. A series of plasma blobs moved 
along the sheet continuously (Figure \ref{fig:blob1}), and five of them 
were well characterized, allowing us to further study important 
properties of the current sheet [{\it Lin et al.} 2007]. Later, another 
event that developed similar morphological features in the disrupting 
magnetic field was observed (Figure \ref{fig:blob2}), {\it Lin et al.} 
[2005] identified five well observed blobs, and {\it Riley et al.} [2007] 
recognized four.

In the numerical experiments of {\it Forbes and Malherbe} [1991], {\it Y. 
Fan} [2005, private communication], and {\it Riley et al.} [2007], the 
repeated formation of a set of blobs that move both toward and away from 
the Sun occurred. Although these numerical experiments were not performed to 
model any specific event, there is good general agreement in the 
formation of the current sheets and the formation and propagation of the 
blobs flowing along the sheet. {\it Forbes and Malherbe} [1991] and 
{\it Riley et al.} [2007] pointed out that the formation and evolution of 
the blob in the sheets are strongly suggestive of the tearing mode 
instability [e.g., {\it Furth et al.} 1963] that plays an important role 
in magnetic field diffusion and governing the scale of the sheet [{\it 
Strauss} 1988; {\it Drake et al.} 2006]. More investigations indicate 
that the turbulent diffusion caused by the tearing mode could be much 
faster or more efficient than that caused by the classical and anomalous 
resistivities [{\it Strauss} 1988; {\it Bhattacharjee and Yuan} 1995].

Alternatively, the formation and propagation of plasma blobs may be 
explained within the framework of unsteady Petschek-type reconnection.
This approach has been employed in the modeling of reconnection in the 
terrestrial context [e.g., {\it Biernat et al.} 1987; {\it Seon et al.} 
1996; {\it Eriksson et al.} 2004]. Because at present {\it in situ} 
measurements are not available to document solar activities, we are not able 
to confirm whether slow shocks would develop during the eruption. 
But the heating and evaporation of the chromosphere in the flare process 
may constitute the indirect evidence of the formation and propagation of 
the slow mode shocks [e.g., see {\it Forbes and Acton} 1996], and our 
knowledge about the reconnection process in the magnetotail also suggests 
the likelihood of the slow mode shocks forming in CME/flare current sheets.

The reconnection outflow along the CME/flare current sheet discussed 
above also shows similar flowing features to those occurring in the 
geomagnetic tail (e.g., see {\it Slavin et al.} [2003 and 2005]). These 
features include both sunward and anti-sunward flows. The sunward flow was 
usually observed to decelerate apparently near the top of the flare loop 
system [{\it Sheeley et al.} 2004; {\it Asai et al.} 2004] as a result of 
deflecting and ceasing by the closed flare loops (e.g., see also {\it 
Reeves et al.} [2008]). Observations indicate as well that the two 
reconnection outflows separate at one of multiple X-points (the joint points 
of every two adjacent plasma blobs or plasmoids) in the sheet (cf. plasma 
flows observed by {\it McKenzie and Hudson} [1999], {\it Sheeley et al.} 
[2004], and {\it Asai et al.} [2004] with those observed by {\it Ko et 
al.} [2003] and {\it Lin et al.} [2005], as well as the numerical 
experiment by {\it Riley et al.} [2007]). This special X-point becomes 
dominant in the eruption and usually does not move with the reconnection 
flow. Similar to the case in the magnetotail, we do not well understand for 
the time being the physical mechanism that determines the location of 
this dominant X-point. But the above observations and numerical 
experiments suggest that it be somewhere between 0.5 and 1.0 
$R_{\odot}$ from the surface of the Sun, and that it is probably governed 
by the nearby plasma and magnetic environment.

\subsection{Mass Flows of Various Scales in the Solar Wind}
Mass flows of various length scales occur ubiquitously in the solar wind. 
Among these flows, the largest feature is the ICME, which is the most 
important product of solar activity in the interplanetary medium, and is 
produced by a rapid release of magnetic energy in the solar eruptive 
process [{\it Lin et al.} 2004 and references therein]. An important 
subset of these is the magnetic clouds [{\it Burlaga et al.} 1981; {\it 
Klein and Burlaga} 1982], which are observed at 1 AU as mesoscale 
(fraction of an AU) configurations characterized by a magnetic field of 
above-average strength executing a large and smooth rotation in a plasma 
of low proton temperature and $\beta$. Their magnetic field structure has 
been modeled as cylindrically-symmetric, constant-$\alpha$ force-free 
magnetic flux ropes of circular cross-section, a  solution of which was 
given by {\it Lundquist} [1950], and has been applied in an 
interplanetary context by {\it Burlaga} [1988], {\it Lepping et al.} 
[1991], and many others. The feet of magnetic clouds may still be 
anchored at the Sun, and examples of such events were reported by {\it 
Farrugia et al.} [1993b and 1993c] and {\it Shodhun et al.} [2000].

Smaller scale flux ropes in the solar wind during quiet times were also 
reported by {\it Moldwin et al.} [2000]. Their average diameter is of 
order a few percent that of magnetic clouds, but in all other aspects 
they are similar to clouds. As generating cause, we suggest reconnection 
in the heliospheric current sheet (as opposed to the corona for magnetic 
clouds). Below we shall discuss small-scale reconnection in the heliospheric
current sheet related to the papers of {\it Gosling et al.} [2007 and 
references therein] and co-workers.

There is a great deal of evidence from both UVCS/SOHO observations and 
{\it in situ} measurements that magnetic reconnection continues
to occur as the solar wind accelerates and expands throughout the
heliosphere [{\it Kohl et al.} 2006; {\it Gosling et al.} 2007]. The 
island-like features are frequently observed in the slow solar wind, 
which is believed to originate (at least in part) from bright ``helmet 
streamers'' that have been seen in white-light coronagraph images for 
decades. But it is uncertain how the plasma expands into a roughly 
time-steady flow since many of these streamers appear to have a closed 
magnetic field. Outflow speeds measured by the UVCS instrument on SOHO 
appear to be consistent with the {\em in situ} slow wind along the 
open-field edges (or ``legs'') of large streamers [{\it Strachan et al.}  
2002]. The closed-field ``core'' regions of streamers, on the other hand, 
exhibit negligible outflow speeds and anomalously low ion abundances 
[e.g., {\it Raymond et al.}  1997]. These observations provide strong 
evidence for the idea that the majority of the slow wind originates along 
the open-field edges of streamers.

However, evidence also exists for a time-variable plasmoid-like component 
of the slow wind in streamers. The increased photon sensitivity of the 
LASCO/SOHO white-light coronagraph over earlier instruments revealed an 
almost continual release of low-contrast density inhomogeneities, or 
``blobs,'' from the cusps of streamers [{\it Sheeley et al.}  1997]. 
These features appear to accelerate to velocities of order 300 to 400 km 
s$^{-1}$ by the time they reach $\sim$30 $R_{\odot}$, the outer limit of 
LASCO's field of view. Because of their low contrast, though (i.e., they 
represent only about 10\% changes from the time-averaged streamer), the 
blobs probably do not comprise a large fraction of the mass flux of the 
slow solar wind.

{\it Wang et al.} [2000] reviewed three general scenarios for the 
production of these blobs: (1) ``streamer evaporation'' as the loop-tops 
(with plasma $\beta \approx 1$) are heated to the point where magnetic 
tension is overcome by high gas pressure; (2) plasmoid formation as the 
distended streamer cusp pinches off the gas above an X-type neutral point; 
and (3) reconnection between one leg of the streamer and an adjacent open 
field line, transferring some of the trapped plasma from the former to the 
latter and allowing it to escape [see Figure 8 of {\it Wang et al.} 2000]. 
They concluded that all three mechanisms might be acting simultaneously 
[see also {\it Wu et al.}  2000; {\it Fisk and Schwadron} 2001; {\it 
Lapenta and Knoll} 2005]. {\it Einaudi et al.} [1999; 2001] performed 
hydrodynamical simulations of the narrow shear layer above the streamer 
cusp and found that magnetic islands can form naturally as the nonlinear 
development of a tearing-mode instability.

There may be some similarity between these scenarios and older models of 
diamagnetic acceleration of the solar wind via buoyant plasmoids 
that may fill some fraction of the corona [e.g., {\it Schl\"{u}ter} 1957; 
{\it Pneuman} 1986; {\it Mullan} 1990]. The additional momentum deposited 
by this process is given approximately by an effective buoyancy-driven
pressure gradient
\begin{equation}
  a_{d} \, = \, - \frac{3}{2} f_{d} w_{d}^{2} \,
  \frac{\partial}{\partial r}
  \left( \ln \frac{B_{0}^2}{8 \pi} \right) \,\,\, ,
\end{equation}
where $f_d$ is the ratio of mass flux in the plasmoids to the total wind 
mass flux and $w_d$ is the most-probable speed in the plasmoids [see also 
{\it Pneuman} 1983; {\it Yang and Schunk} 1989; {\it Tamano} 1991]. 
Plasmoid inhomogeneities may arise from reconnection events [e.g., 
{\it Yokoyama and Shibata} 1996] and expand to fill a large fraction of the 
volume of coronal holes [see also {\it Feldman et al.}  1997].

At some point in the outer solar atmosphere, MHD turbulence develops
from the underlying waves and/or field-line motions (e.g., see
{\it Coleman} [1968]; {\it Goldstein et al.} [1995]; {\it Tu and Marsch} 
[1995]). The energy flux in the turbulent cascade has been proposed as a
natural means of heating the corona and accelerating the solar wind [{\it 
Hollweg} 1986; {\it Matthaeus et al.} 1999; {\it Cranmer et al.} 2007].
Interestingly, on the smallest spatial scales, MHD turbulence has been 
shown to develop into a collection of narrow current sheets undergoing 
oblique magnetic reconnection (i.e., with the strong ``guide field'' 
remaining relatively unchanged). There is increasing evidence for the 
presence of such small-scale reconnection regions in both the slow solar 
wind [{\it Gosling et al.} 2005, 2006] and the fast solar wind [{\it 
Gosling et al.} 2007], both of which are strongly turbulent. These {\em 
in situ} measurements generally pinpoint regions of fast jetlike 
``exhaust'' bounded by pairs of back-to-back rotational discontinuities 
in solar wind streams having low plasma beta ($\beta < 1$). Reconnection 
in these regions has been suggested as a possibly important channel for 
maintaining the extended heating of the solar wind (e.g., {\it Leamon et 
al.} [2000]).

Small-scale current sheets and plasmoids arise spontaneously in numerical 
simulations of MHD turbulence.  These simulations are generally not seeded 
with any kind of forced reconnection. The coherent structures arise from 
the natural stochastic evolution of the fields (e.g., see {\it Kinney et 
al.} [1995]; {\it Mininni et al.} [2006]). {\it Dmitruk et al.}  [2004] 
and {\it Dmitruk and Matthaeus} [2006] performed test-particle simulations 
in a turbulent plasma that contains the kinds of reconnection regions 
described above. They found that protons can become perpendicularly 
accelerated around the guide field because of coherent forcing from the 
perturbed fields associated with the current sheets. Also, {\it Markovskii 
et al.} [2006] found that the intensified shear motions across these 
current sheets can lead to an unstable growth of waves near harmonics of 
the ion-cyclotron frequency.  These ideas may be important ingredients in 
the production of preferential heating and acceleration of heavy ions in 
the extended solar corona, as observed by UVCS/SOHO (e.g., see {\it Kohl 
et al.} [1997 and 2006]).

\section{Mechanisms for Producing Plasmoids (Plasma Blobs)}
Small scale structures turn out to be ubiquitous in reconnecting current 
sheets in various circumstances. Their existence is essential for 
improving the efficiency of energy conversion through magnetic 
reconnection, and thus for enhancing the rate of reconnection. In the 
framework of plasma physics, these structures are ascribed to plasma 
instabilities and the corresponding modes of turbulence; and in the 
framework of MHD, on the other hand, they could be the reconnection 
outflow regions surrounded with the slow mode shocks. Here, we discuss 
these two scenarios and their implications for observations. It is 
difficult to distinguish between them on the observational evidence we 
have at present.


%
%

\subsection{Current Sheets Undergoing Tearing}
Traditional Sweet-Parker reconnection [{\it Sweet} 1958a and 1958b; 
{\it Parker} 1957 and 1963] is too slow to account for the rate of energy 
conversion occurring in either the geomagnetic substorm or the solar 
eruption. The low rate of energy conversion results from the large scale 
of the diffusion region. Reducing such a scale helps the energy conversion
become more efficient. Minimizing the diffusion region scale in a large  
(long) current sheet could be well accomplished by the plasma turbulence
invoked by various instabilities (e.g., {\it Furth et al.} [1963]; {\it 
Ambrosiano et al.} [1988]; {\it Lazarian and Vishniac} [1999]; {\it Priest 
and Forbes} [2000]; {\it Drake et al.} [2006]). In this case, the scale of 
the diffusion region is roughly that of the turbulence eddies, or magnetic 
islands, which is much smaller than the global length scale of the system, 
and yields higher reconnection rate [{\it Strauss} 1988; {\it 
Bhattacharjee and Yuan} 1995]. The reconnecting current sheet in such a 
framework is an assembly of many modes of turbulence or the filamental 
current sheets and reconnection exhausts, which is the typical scenario of 
turbulence reconnection.

\begin{figure}
\noindent\includegraphics[width=29pc]{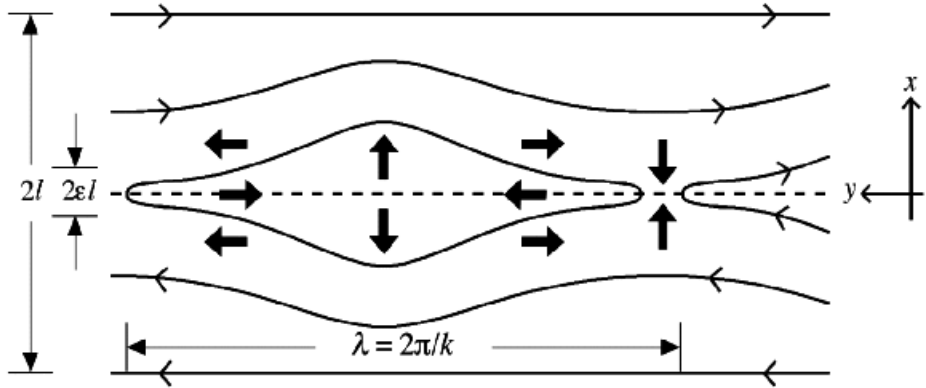}
\caption{Interior structure of the current sheet in which the tearing   
mode instability develops. Thick arrows show plasma flow and thin arrows 
are for magnetic field lines. (Courtesy of E. R. Priest.)}   
\label{fig:tearing}
\end{figure}

Among these instabilities, the tearing mode is one of the most 
extensively studied. It is a long-wavelength, resistive instability. It 
was investigated for the first time by {\it Furth et al.} [1963] in the 
framework of resistive instability modes established by {\it Dungey} 
[1958], which showed that at an X-type neutral point configuration in the 
magnetized plasma, finite conductivity (or resistivity) can give rise to 
an unstably growing current concentration. By this mechanism, a sheet 
pinch, or a current sheet, can tear along current-flow lines, forming a 
chain of filaments, or magnetic islands in projection (Figure 
\ref{fig:tearing}). Here $k$ is the wave number of the turbulence caused 
by the instability and $l=d/2$ is the half thickness of the sheet.

Properties of the magnetized plasma require that the growth rate of the
modes considered be slow compared to the hydromagnetic rate, but fast    
compared to the resistive diffusion rate. In the case of the tearing    
mode, this relates $k$ to $l$ by [{\it Furth et al.} 1963]
\begin{equation}
S^{-1/4}<kl<1, \label{eq:ineq}
\end{equation}
where $S=\tau_{d}/\tau_{A}$ is the Lundquist number of the current sheet,
$\tau_{A}=l/V_{A}$ and $\tau_{d}=l^{2}/\eta$ are the times at which
the Alfv\'{e}n wave and the resistive diffusion traverse the sheet,
respectively. Here, $V_{A}$ is the local Alfv\'{e}n speed near the current
sheet, and $\eta$ is the magnetic diffusivity of the sheet. The turbulence
caused by the instability broadens the current sheet significantly so that
$S\gg 1$ usually holds in most circumstances. From the definition of $S$,
we have $S=lV_{A}/\eta$.

According to the standard theory of magnetic reconnection, the magnetic
field is continuously being dissipated through the current sheet at the 
same rate as magnetic flux is brought into the sheet, which leads to [e.g., 
see {\it Priest and Forbes} 2000, p. 120]:
\begin{equation}
v_{i} = \frac{\eta}{l}, \mbox{ with } \eta =
\frac{\eta_{e}}{\mu_{0}}, \label{eq:diffu}
\end{equation}
where $v_{i}$ is the reconnection inflow speed in units of m s$^{-1}$,  
$\eta$ is in units of m$^{2}$ s$^{-1}$, $\eta_{e}$ is the electric 
resistivity in ohm m, and $\mu_{0}=4\pi\times 10^{-7}$ H m$^{-1}$. Then we
end up with
\begin{equation}
S=V_{A}/v_{i}=M_{A}^{-1}, \label{eq:SMa}
\end{equation}
where $M_{A}$ is the Alfv\'{e}n Mach number, the rate of magnetic
reconnection measured as $v_{i}/V_{A}$, which is also known as the 
relative reconnection rate compared to the absolute reconnection rate 
that is measured as the reconnecting electric field in the current sheet 
(see discussions by {\it Lin and Forbes} [2000] and {\it Priest and 
Forbes} [2000], on this issue). Equation (\ref{eq:SMa}) relates $S$ 
directly to $M_{A}$, a quantity that can be deduced from observations.

After going through simple algebra, we find from $kl > S^{-1/4}$ of
(\ref{eq:ineq}):
\begin{equation}
l_{min}=k^{-1}S^{-1/4}=M_{A}^{1/4}\frac{\lambda}{2\pi}, \label{eq:l_min}
\end{equation}
where $\lambda$ is the wavelength of the turbulence (see Figure
\ref{fig:tearing}), and is identified with the distance of two successive
plasma blobs flowing along the current sheet [see {\it Ko et al.} 2003; 
{\it Lin et al.} 2005]. Therefore, equation (\ref{eq:l_min}) relates 
$l_{min}$ to $M_{A}$ and $\lambda$ in a simple and straightforward 
fashion. Usually, $M_{A}$ in reality varies from $10^{-3}$ to $10^{-1}$ 
for solar events and laboratory plasma [e.g., {\it Furth et al.} 1963; 
{\it Priest and Forbes} 2000; {\it Yokoyama et al.} 2001; {\it Ko et al.} 
2003; {\it Lin et al.} 2005], and from $10^{-2}$ to $10^{-1}$ for the 
reconnection processes in both the magnetopause [{\it Vaivads et al.} 2004; 
{\it Rentin\`{o} et al.} 2007; references therein) and the magnetotail 
[{\it Xiao et al.} 2007; references therein]. Thus in these disparate 
contexts, $M_{A}$ has approximately the same range of variation.

These values of $M_{A}$ indicate that $S\gg 1$ usually holds for both
solar and geomagnetic cases, but $S^{1/4}\gg 1$ does not. Instead, 
$S^{1/4}>1$ holds marginally for most of these values of $M_{A}$. Hence, 
by inequalities in (\ref{eq:ineq}), quantity $kl$ could possess a finite 
lower limit. Direct observations of the CME/flare current sheets indicate 
that the sheet thickness varies from $10^{4}$ to $10^{5}$ km [{\it Lin et 
al.} 2007], and {\it in situ} measurements of the magnetopause and the 
magnetotail estimate a thickness ranging from 200 to 500 km [{\it 
Vaivads et al.} 2004; {\it Xiao et al.}  2007].

Because of the size of the sheet thickness and the energy of the electrons 
there contained, collisional effects (such as Spitzer resistivity) cannot 
play a role in the dynamics of the current sheet. Instead, the relevant 
processes are collisionless with macroscopic effects that transform 
magnetic energy into plasma kinetic and thermal energy, and could account 
for various timescales of the phenomena during auroral events in the 
auroral regions and in the magnetotail [{\it Coppi 1965}]. However, 
subsequent investigations on the tearing mode in the magnetotail [{\it 
Pritchett and Coroniti} 1990; {\it Pritchett et al.} 1991] revealed that 
the ion tearing mode is unlikely to develop spontaneously in the closed 
field line region of the near-Earth plasma sheet, but external influences 
and time-dependent effects (with the attempt to establish large-scale 
convection flow) might be able to trigger the tearing mode instability and 
lead to rapid reconnection.

Collisionless effects become dominant when the resistive scale size is 
smaller than the ion skin depth, $\sqrt{m_{i}c^{2}/(4\pi n_{i} e^{2})}$, 
where $m_{i}$ is the ion mass, $n_{i}$ is the number density, $e$ is the 
charge. In this case a separation of ions and electrons results. The
ions decouple from the magnetic field in the ion diffusion region.
Eventually, inside the smaller electron diffusion region,
the electrons, too, become demagnetized [{\it Vasyliunas} 1974]. The
separation of ions and electrons in the ion diffusion region
produces a system of currents (Hall currents) which induce a quadrupolar 
Hall magnetic field pattern [{\it Sonnerup} 1979]. This outlines a 
scenario of collisionless tearing mode turbulence and its observational 
consequences, and the formation of Hall magnetic and electric field 
structure in the vicinity of the diffusion region is a key feature of 
collisionless magnetic reconnection.

Investigations of collisionless reconnection by means of particle 
simulations were reported in the context of the geospace environmental 
modeling reconnection challenge by {\it Birn et al.} [2001]. Another 
example of studying collisionless reconnection via particle-in-cell 
simulations was conducted recently by {\it Shay et al.} [2007]. 
They demonstrated that reconnection remains fast in very large 
systems. The electron dissipation region develops a distinct two-scale 
structure along the outflow direction. Consistent with fast reconnection, 
the length of the electron current layer stabilizes and decreases with 
decreasing electron mass, approaching the ion inertial length for a 
proton-electron plasma. They found that the electrons form a 
super-Alfv\'{e}nic outflow jet that remains decoupled from the magnetic 
field and extends large distances downstream from the X-line.

In the magnetotail context, a number of spacecraft observations have been 
reported where the quadrupolar Hall magnetic fields were detected.
When a spacecraft traverses the magnetotail reconnection region from one 
side of the X-line to the other, it could confirm the presence of the Hall
quadrupolar fields, and by implication the importance of collisionless 
effects in reconnection, if it observed a reversal of the out-of-plane 
(i.e. parallel to the X-line) magnetic field component [{\it Shay et al.} 
2001]. The Hall quadrupolar magnetic fields have been observed by the 
Geotail spacecraft in the near-Earth region [{\it Nagai et al.} 2001] 
outside the diffusion region. {\it In situ} detection of signatures of 
Hall currents were presented by {\it {\O}ieroset et al.} [2001], who 
analyzed data returned by the Wind spacecraft as it traversed the near - 
Earth ($\sim$ 60 $R_{E}$) geomagnetic tail going from the earthward to 
tailward side of the X-line.

In the solar context, considering plasma properties of the coronal 
environment, magnetic reconnection occurring in the corona is 
collisionless as well. It has now been suggested that the collisionless 
aspects of reconnection are important to regulate the rate of coronal 
heating (e.g., see recent works by {\it Cassak et al.} [2006 and 2008]; 
{\it Uzdensky} [2007]), in addition to that releasing the magnetic 
energy in the CME/flare current sheet as shown in Figure 
\ref{fig:CME-flare}.

Recent multi-point Cluster observations were conducted by {\it Eastwood et 
al.} [2007]. They observed a reconnection event in the near-Earth 
magnetotail where the diffusion region was nested by the Cluster 
spacecraft. They found that, close to the diffusion region, the magnetic 
field displays a symmetric quadrupole structure, the Hall electric field 
is strong (up to $\sim 40$ mV m$^{-1}$) on the earthward side of the 
diffusion region, but substantially weaker on the tailward side. In 
conjunction with these observations, a small-scale magnetic flux rope was 
observed. It was located in the plasma flow near the reconnection site and 
may be the secondary islands appearing in numerical experiments. A very 
strong Hall electric field (up to $\sim 100$ mV m$^{-1}$) inside the 
magnetic island was detected.

\subsection{Other Types of Multiple X-Line Reconnection}
In addition to the tearing mode instability and the consequent turbulence, 
other possibilities exist responsible for the magnetic island formation in 
the reconnecting current sheets. Among these possibilities is the 
Petschek reconnection, which was first investigated by {\it Petschek} 
[1964] for the steady-state case. Considering the time-dependent behaviors 
of the process in both solar flares and substorms in the magnetosphere, 
Petschek's original model was developed to the time-dependent version 
(e.g., see {\it Pudovkin and Semenov} [1985]; {\it Biernat et al.} 
[1987]; {\it Priest and Forbes} [2000], pp. 222--229).

In this framework, the classical Ohmic diffusion is confined to a small 
region, and disturbance caused by the diffusion generates the slow mode 
shocks propagating rapidly into the surrounding plasma medium, and 
transmitting the diffusion-associated disturbances into the system at 
large [e.g., see {\it Biernat et al.} 1987]. There is a self-adjustment 
between the large scale field and plasma flow distribution and the 
reconnection rate, so the evolutionary behavior of the hydromagnetic 
configuration is ultimately governed both by the globally imposed boundary 
conditions and the local processes occurring inside the diffusion region, 
which is described by the time-dependent Alfv\'{e}n Mach number 
$M_{A}(t)$.

\begin{figure}
\noindent\includegraphics[width=26pc]{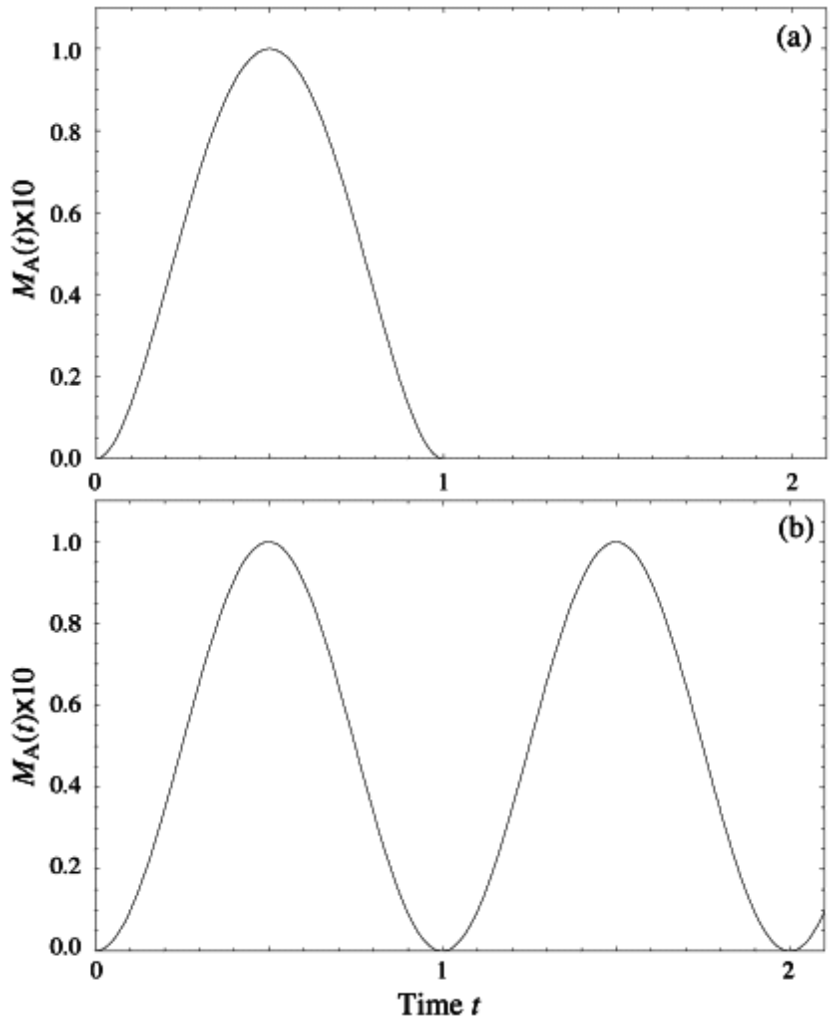}
\caption{Variations of $M_{A}(t)$ versus time. ($a$) Single pulse
reconnection governed by (\ref{eq:MA_SP}), and ($b$) bursty reconnection
described by (\ref{eq:MA_MP})}
\label{fig:MA}
\end{figure}

As examples, we study the response of the medium to the following two time 
profiles of $M_{A}$:
\begin{eqnarray}
M_{A}(t)&=&\frac{5}{8}t^{2}(t-1)^{2} \hspace{5mm} \mbox{for } 0<t<1
\nonumber\\
&=& 0 \hspace{5mm} \mbox{otherwise}, \label{eq:MA_SP}
\end{eqnarray}
and
\begin{equation}
M_{A}(t)=\frac{1}{20}[1-\cos(2\pi t)], \label{eq:MA_MP}
\end{equation}
where $t$ is in units of $\tau_{A}$. Time profiles of these two $M_{A}$s 
are plot in Figures \ref{fig:MA}$a$ and \ref{fig:MA}$b$, respectively.
Magnetic reconnection described by (\ref{eq:MA_SP}) commences at $t=0$, 
reaches its maximum 0.1 at $t=0.5$ and ceases at $t=1$.

\begin{figure}
\noindent\includegraphics[width=25pc]{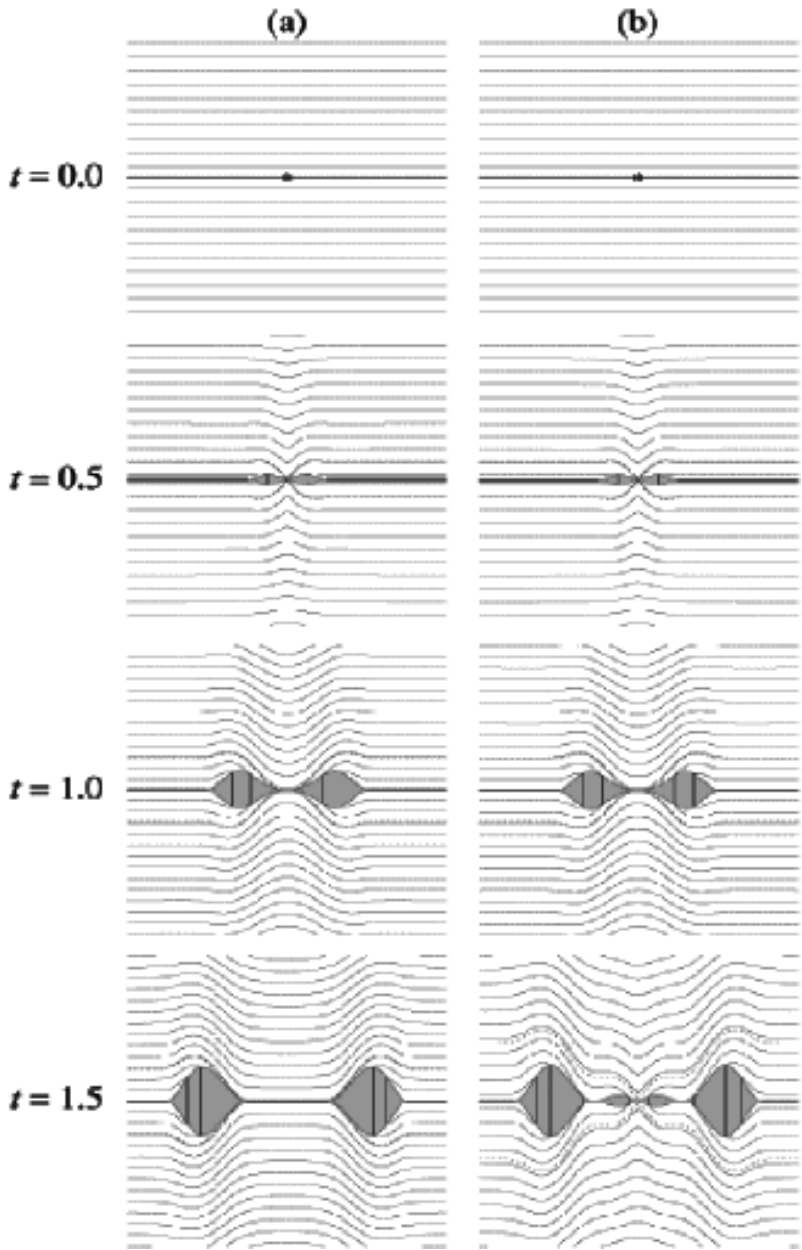}
\caption{Evolution of the hydromagnetic configurations in the time-dependent
Petschek-type reconnection in response to the single pulse dissipation
($a$), and to the bursty dissipation ($b$), respectively. Panels are for
the snapshots of the hydromagnetic configurations at different times. The
asterisks in the panels for $t=0$ indicate the location where
reconnection is initiated, solid curves describe magnetic field lines,
thick solid lines are for the current sheets, dashed curves manifest the
separatrices, and the shadowed areas are the reconnection outflow regions
surrounded by the slow mode shocks. The $x$-axis points to the right,
the $y$-axis points upward, and the origin is co-located with the
asterisk. The scale in $y$-direction in each panel has been enlarged by a
factor of 10 in order to display detailed structures of the disturbance.}
\label{fig:TD_Petschek}
\end{figure}

In the case of single pulse reconnection, the response of the medium is 
displayed in Figure \ref{fig:TD_Petschek}$a$. The disturbance commences at 
the origin (namely the diffusion region marked with asterisk) as $t=0$, 
and propagates at the local Alfv\'{e}n speed, which is larger than the 
slow magnetoacoustic speed, so the slow mode shock forms in front of the 
propagating disturbance ($t=0.5$). Collectively, these slow shocks 
establish an outflow region for plasma and magnetic field streaming along 
the current sheet. Upstream of the shock, the medium is unperturbed due to 
the finite propagating speed of the disturbance. With reconnection ceasing 
after $t=1.0$, no further perturbation is produced from the diffusion 
region, the outflow region and the slow shocks are detached from the 
original diffusion region, the current sheet is re-established in the 
wake of the outflow region ($t\ge 1.0$), and the reconnection inflow is 
prevented but the outflow regions keeps propagating and expanding. (Note: 
Detailed formulations for magnetic configurations in this figure and 
in Figure \ref{fig:TD_Petschek}$b$ below can be found in {\it Biernat et 
al.} [1987] and /or {\it Priest and Forbes} [2000, pp. 222--239]. We do 
not duplicate them here.)

Magnetic reconnection described by (\ref{eq:MA_MP}) produces a different 
evolution in the system. It commences at $t=0$, but does not cease. So a 
series of reconnection outflow regions associated with slow shocks are 
continuously formed and move away from the diffusion region (Figure 
\ref{fig:TD_Petschek}$b$). This corresponds to a bursty reconnection 
scenario in which plasma blobs or plasmoids are observed to develop 
successively (see also {\it Riley et al.} [2007] for examples of 
numerical experiments). It represents some important features of the 
multiple X-line reconnection process as well. We note that we use the 
single pulse reconnection here as an example to demonstrate how the impact 
of the magnetic diffusion occurring in a local region could cause 
disturbance and energy conversion in the medium at large. But it may not 
be the fashion in which magnetic reconnection takes place in reality. 
Instead, bursty reconnection is more likely to occur in both solar and 
magnetospheric environments.

The most significant consequence of introducing the slow mode shocks 
attached to the diffusion region is the tremendous enhancement of the 
energy conversion rate, or the rate of magnetic reconnection since the 
slow mode shock is also known as a switch-off shock through which the 
downstream tangential magnetic field is almost switched off, and the 
corresponding magnetic energy is converted into heating and kinetic energy 
of the reconnected plasma [{\it Priest} 1982]. The density, the pressure, 
and the temperature of the reconnected plasma in the outflow region 
increase by a factor of
\begin{displaymath}
r_{n}=\frac{\gamma(\beta+1)}{\gamma(\beta+1)-1},
\hspace{0.25cm} r_{p}=\frac{\beta+1}{\beta}, \hspace{0.25cm}\mbox{and}
\end{displaymath}
\begin{equation}
r_{T}=\frac{\beta+1}{\beta} \frac{\gamma(\beta+1)-1}{\gamma(\beta+1)}
\label{eq:ratio}
\end{equation}
respectively, with $\gamma$ being the ratio of specific heats and $\beta$
being the plasma beta (e.g., see {\it Pudovkin and Semenov} [1985]). 
Usually, $\gamma = 5/3$, so equations in (\ref{eq:ratio}) implies 
significant heating when reconnection occurs in the force-free 
($\beta\ll 1$) environment like the solar corona and the magnetotail 
lobes, very weak heating in the large $\beta$ environment like the 
photosphere and the dayside magnetopause (refer to Table \ref{tbl:VA}).

In the solar corona, the consequences of the bursty reconnection are 
identified in several eruptive events that have been well studied (see 
Figures \ref{fig:tadpole}, \ref{fig:blob1}, and \ref{fig:blob2}), and the 
slow mode shock may also account for observations of high temperature 
emissions [{\it Innes et al.} 2001; {\it Ciaravella et al.} 2002; 
{\it Ko et al.} 2003; {\it Raymond et al.} 2003]. In the Earth's 
magnetotail, such a reconnection mode has been widely held to be a
mechanism by which plasmoids/magnetic flux ropes are produced. This view
is grounded on analytical modeling (e.g. {\it Schindler} [1974]), and  
extensive numerical simulations [{\it Hesse and Birn} 1991; {\it Hesse
et al.} 1996; {\it Hesse and Kivelson} 1998; and references therein]. 
There is also the work of {\it Fu and Lee} [1985] on multiple X-line 
formation that will be discussed shortly. The events associated with the 
slow mode shocks have been reported as well (e.g., see {\it Seon et al.} 
[1996]; {\it Eriksson et al.} [2004]; and also {\it Priest and Forbes} 
[2000], pp. 343--344 for a brief overview). A reconnection layer in the 
solar wind where slow mode shocks were tentatively identified was given by 
{\it Farrugia et al.} [2001].

Scaling laws associated with the formation of the slow mode shocks 
connect the current sheet thickness to the extension of the slow mode 
shocks, and can be deduced from the shape or structure of the slow mode 
shocks. From equations (11) and (17c) of {\it Biernat et al.} [1987], or 
equivalently equation (7.47) of {\it Priest and Forbes} [2000], we find 
the average sheet thickness $\bar{l}$ can be related to the average 
extension of the shock $\bar{\lambda}$ via the average rate of magnetic 
reconnection $\bar{M_{A}}$
\begin{equation}
\bar{l}\approx r_{n}^{-1}\bar{M_{A}}\bar{\lambda}, \label{eq:average}
\end{equation}
in the Petschek reconnection process. Here $\bar{M_{A}}$ and 
$\bar{\lambda}$ are obtained by taking averages of $M_{A}(t)$ over the 
time domain and of the shock extension over the space domain in 
$x$-direction (see Figure \ref{fig:TD_Petschek}), respectively, with 
keeping the fact in mind that the medium upstream of the shock remains 
unperturbed before the shock arrives. Forms of those equations used to 
deduce \ref{eq:average} are simple and the algebra performed are 
straightforward, so we do not duplicate them here and skip the algebra. 
Similar to (\ref{eq:l_min}), equation (\ref{eq:average}) relates several 
important parameters for magnetic reconnection to one another, and these 
parameters are all observables. Such relations provide us opportunities to 
look into some crucial properties of energy conversion through magnetic 
reconnection in either the magnetosphere or the solar corona, and may help 
tell what type of reconnection is occurring or has occurred by comparing 
theories and observations.

We notice the similarity of equation (\ref{eq:average}) to
(\ref{eq:l_min}) in that $l$ is related to $\lambda$ linearly although
these two equations are deduced in different frameworks. Such as
similarity is quite likely to suggest a kind of equivalence in physics   
of the two frameworks in which the reconnection process is studied. This
is probably because both turbulent eddies and shocks act as diffusion
structures which play an important role in converting energy from one
form into another. If we enlarge the fine structures of the turbulent   
current sheet, we may find that each magnetic island-neutral point-magnetic
island structure inside the turbulent current sheet in Figure
\ref{fig:tearing} can be seen as a Petschek-type reconnection morphology
displayed in Figure \ref{fig:TD_Petschek} in small scale, and a turbulent
sheet should be full of numerous such small scale structures [see {\it Drake
et al.} 2006; {\it Rentin\`{o} et al.} 2007].

In fact, both the turbulence and the Petschek versions of reconnection
were used to explain the same results of a numerical experiment. In
their original work on the numerical simulation of magnetic reconnection
occurring in the current sheet extending from the top of the flare loop 
system, {\it Forbes and Malherbe} [1991] identified the plasma blobs 
appearing in the sheet with the magnetic islands caused by the tearing 
mode. But {\it Priest and Forbes} [2000, pp. 391$-$393] explained these 
blobs within the framework of the Petschek-type reconnection. That 
alternative versions of reconnection were used to account for the same 
result of numerical experiment should somehow imply the viability of 
the two approaches to investigating and describing the reconnection 
process. Furthermore, this might suggest that both versions of 
reconnection operate together in the eruptive event in reality. At 
present this issue requires further investigations.

In addition, multiple X-line reconnection may occur in a different 
fashion. Numerical investigations of the plasma dynamics of a long 
magnetotail ($\sim 200 R_{E}$) indicated that the formation of X-line and 
plasmoids occurs intermittently and repeatedly every 2 -- 4 hours [{\it 
Lee et al.} 1985], and reconnection in the magnetotail during substorms 
and storms is basically a driven process. These results are consistent 
with some observations made in the magnetotail. The force driving 
reconnection is governed by the environment and manifests time-dependent 
features, which may account for the time profile of $M_{A}$ discussed 
above. The formation and evolution of a three-dimensional plasmoid in the 
geomagnetic tail was numerically studied by {\it Hesse and Birn} [1991]. 
The configuration investigated includes the transition from a closed 
field line region to an open far-tail region with a distant X-line. They 
found that the formation and growth of a plasmoid body through 
reconnection occurs in the near-Earth region, and helical plasmoid field 
lines are produced but remain connected to the Earth.

Besides the phenomena in the magnetotail, the multiple X-line reconnection 
may also result in FTE's observed by ISEE satellites at the dayside 
magnetopause [{\it Lee and Fu} 1985]. Existence of the skew components of 
the reconnected field lines allows the formation of the flux rope 
consisting of the helical field lines. In this model the magnetic island 
is not isolated but is part of an interconnected chain of two-dimensional 
X-points and O-points. Further studies showed that the occurrence of the 
multiple X-line reconnection is governed by the current sheet scales 
[{\it Fu and Lee} 1985]. This process is expected to happen when the sheet 
length is large compared to its thickness and the resistivity is low. 
{\it Lee and Fu} [1986] found as the ratio $\lambda/l$ exceeds a critical 
value between 10 and 20, multiple X-lines are observed to appear in the 
current sheet.

{\it Lee} [1990] noted that with the formation of a long current sheet in 
the magnetopause, the non-uniformity of the solar wind pressure and the 
stress of interplanetary magnetic flux exerted on the magnetopause surface 
may also cause variations in the current density and, consequently, 
reconnection to occur at several locations. Associated with the 
successive appearance of multiple X-lines and plasmoids is bursty 
reconnection rate or electric field [{\it Lee et al.} 1990], which 
resembles that shown in Figure \ref{fig:MA}$b$. As we already mentioned 
before, features of multiple X-line reconnection have also been recognized 
in both magnetotail substorms [{\it Slavin et al.} 2003 and 2005] and solar 
eruptive events [{\it Ko et al.} 2003; {\it Lin et al.} 2005], as well as 
in the numerical experiments for solar eruptions [{\it Riley et al.} 
2007].

Before ending this section, we believe it is also worth mentioning the 
work by {\it Lazarian and Vishniac} [1999] who studied the role of the 
stochastic features in the current sheet through a different approach. In 
this version of energy conversion, magnetic field dissipation starts with 
Ohmic diffusion. Then, stochastic components of the magnetic field are 
produced in the initial stage. With the occurrence of stochastic 
components, there is small-scale ``wandering" in the field lines. This 
allows magnetic reconnection to take place among adjacent wandering field 
lines (see Figure 2 of {\it Lazarian and Vishniac} [1999]), yielding 
multiple reconnection sites within the sheet. The presence of multiple 
reconnection sites results in a minimum rate of reconnection, 
$M_{A}=R_{L}^{-3/16}$, where $R_{L}$ is the magnetic Reynolds number in 
the whole system involved in the energy conversion. In the solar coronal 
$R_{L}$ ranges from $10^{8}$ to $10^{12}$, which brings the minimum of 
$M_{A}$ to $5.6\times 10^{-3}$ that is within the range of the observed 
values that we have known so far.

\section{Discussion}
In this work we have engaged in a comparative study of the reconnection 
process in the solar and terrestrial contexts. Properties of the 
magnetized plasma determines that the rate of reconnection in both 
environments is controlled by the local Alfv\'{e}n speed. This further 
yields that the reconnection process in the low $V_{A}$ and high $\beta$ 
region transports magnetic flux and energy to the region of high 
$V_{A}$ and low $\beta$ gradually, and that reconnection in the high 
$V_{A}$ and low $\beta$ region releases the stored magnetic energy 
violently. This may explain why the consequences occurring in the 
photosphere and in the dayside magnetopause (low $V_{A}$ and high 
$\beta$) are much less significant than those occurring in the corona 
(flares and CMEs) and in the magnetotail (substorms) (high $V_{A}$ and 
low $\beta$).

Of course, because of the huge difference in scale length and magnetic field
strength, the absolute amount of energy conversion and the absolute rate
of reconnection (usually measured by the strength of reconnection
electric field along the X-line) in the solar and terrestrial contexts 
could greatly differ from one another. For example, a typical solar 
eruption involves energy release of up to 10$^{32}$ ergs (e.g., see {\it 
Priest} [1982]) and the absolute reconnection rate typical varies from 1 
V/cm to 10 V/cm (e.g., see {\it Poletto and Kopp} [1986] and {\it Qiu et 
al.} [2004]), and a typical substorm releases energy of only $10^{21}\sim 
10^{22}$ ergs [{\it Baker et al.} 1997] and the absolute reconnection 
rate of 1 $\sim$ 10 mV/m (e.g., see {\it Blanchard et al.} [1997] and 
{\it Vaivads et al.} [2006]).

On the other hand, the rate of reconnection is also often measured by the 
Alfv\'{e}n Mach number $M_{A}$, which is the reconnection inflow speed 
$v_{i}$ compared to the local Alfv\'{e}n speed $V_{A}$ near the 
reconnection region. So $M_{A}$ is known as the relative reconnection 
rate as well (e.g., see discussions of {\it Lin and Forbes} [2000] and 
{\it Priest and Forbes} [2000]). In this sense, $M_{A}$ is within the 
same order of magnitudes between 0.01 and 0.1 for the reconnection 
processes in both solar and magnetospheric environments (cf. {\it Ko et 
al.} [2003], {\it Lin et al.} [2005] and {\it Vaivads et al.} [2006]) 
although the absolute rates of reconnection in the two environments are 
vastly different. This result should help us study and well understand 
the similar energy conversion processes taking place on other objects in 
the universe.

Another emphasis related to reconnection has been the formation of 
plasmoids/plasma blobs in both environments. The magnetic reconnection 
process manifests almost the same features in various different 
magnetized plasma environments, including the solar corona and the 
magnetosphere. Usually flows are observed in two opposite directions 
along the sheet in both the magnetotail and the CME/flare sheet, and only 
the antisunward flow is observed in the helmet streamer. This is probably 
because the reconnection in the helmet streamer is weaker and the sunward 
reconnection outflow is easily stopped by the closed loops below. 
Plasmoids or plasma blobs flowing along the reconnecting current sheet 
are the ubiquitous features observed in these environments although 
mechanisms for the formation the current sheet vary from case to case.

These plasmoids are usually ascribed to multiple X-line reconnection 
as a result of the tearing mode instability developing in a long thin 
current sheet (e.g., see {\it Furth et al.} [1963] and {\it Priest} 
[1985] for the review; and {\it Drake et al.} [2006] and {\it Loureiro et 
al.} [2007] for the most recent works on this topic). Reconnection 
occurs at each of these X-lines, but one of them eventually develops to 
become the dominant one at which the fastest reconnection takes place, 
and its high speed flow expels both plasmoids and other X-lines to move 
in two opposite directions. The mechanism determining the location of this 
dominant X-line is an open question, but it should depend on the 
parameters of the plasma and magnetic field around.

An alternative process that may account for the plasmoid feature is
time-dependent Petschek-type reconnection extensively studied by Semenov 
and his co-authors. In this framework, diffusion initially commences 
somewhere in a pre-existing current sheet where a finite electrical 
resistivity appears. The impact caused by the diffusion is not confined to 
the diffusion region but propagates as a disturbance at the local 
Alfv\'{e}n speed to the medium at large. Determined by the time profile 
of the diffusion, the disturbance propagates along the current sheet with 
two pairs of the slow mode shocks surrounding the reconnected plasma and 
magnetic field in the outflow region. Departing from the diffusion 
region, the two pairs of the slow shock and the associated plasma flows 
propagate in opposite direction. In this case, multiple X-lines and blobs 
may appear alternatively along the sheet, but the dissipation just occurs 
only at the slow shocks and the X-line where the initial diffusion 
commences. Such a treatment applied by Semenov and his co-workers is 
based on considerations of mathematical simplicity in order to find
analytic solutions to the problem.

This scenario is quite like that of the multiple X-line reconnection due 
to the tearing mode in the stage when one X-line eventually becomes dominant 
and two reconnection outflows in opposite direction emanate from it. One
difference is that the dissipation occurs everywhere in the sheet 
undergoing tearing, and the dominant X-line develops in a self-consistent 
fashion from many X-lines during the process (e.g., see {\it Fu and Lee} 
[1985]). But the dominant X-line in the Petschek-type reconnection is 
fixed artificially from the very beginning. Another difference comes from 
our understanding of the dissipation mechanisms in the process. In the 
tearing mode version, the dissipation occurs in the whole current sheet as 
a result of the plasma turbulence caused by the instability. In the 
Petschek version, on the other hand, in addition to the diffusion region, 
the slow shocks play the main role in the energy conversion. These two 
approaches should be physically equivalent. Comparing the arguments and 
discussions of {\it Forbes and Malherbe} [1991] and {\it Priest and 
Forbes} [2000, pp. 391--393] on the same results from the numerical 
experiment of reconnection in the two-ribbon flare process seems to be 
suggestive of such an equivalence. But rigorous studies of this 
equivalence are not yet available due to mathematical difficulties.

In addition to the interior structure of the reconnecting current sheets, 
it is also necessary to pay attention to the environment outside 
the sheet and its possible impacts on the energy conversion process. In 
discussions of magnetotail versus corona reconnection, one topic that has 
come up repeatedly is whether there is any evidence for the strong 
increases in Alfv\'{e}n speed at the Sun as one moves normal to the 
reconnecting sheet. The gradients are large in the Earth's tail with 
Alfv\'{e}n speeds of $< 100$ km s$^{-1}$ in the central plasma sheet 
where reconnection commences, but increasing to $> 1000$ km s$^{-1}$ over 
100 to 1000 km distant in the outer plasma sheet and lobes. This strong 
gradient is believed to be the source of the ``explosive" appearance of 
magnetotail reconnection as first the low Alfv\'{e}n speed flux tubes in 
the central plasma sheet reconnect followed later by the high Alfv\'{e}n 
speed lobe flux tubes (e.g., see {\it Hesse et al.} [1996]).

In principle, fast reconnection in the solar corona also occurs during 
the explosive event, like flares, in the current sheet where a large 
gradient of Alfv\'{e}n speed exists in the direction normal to the sheet. 
This is represented indirectly by the dramatic change in plasma $\beta$ 
from well below unity outside the sheet to around unity inside the 
sheet (e.g., see {\it Ko et al.} [2003]), and by the significant 
difference between the inflow and the outflow speeds of the plasma 
measured around the sheet developed in an eruption (e.g., see {\it Lin et 
al.} [2005]). Unlike the geomagnetic tail sheet in the ``explosive" stage, 
however, the coronal sheet in the ``explosive" phase does not exist 
beforehand, but forms and develops during the eruptive process. So the 
strong gradient of the Alfv\'{e}n speed across the CME/flare sheet is not 
the very source of the ``explosive" appearance of reconnection. Instead 
the CME/flare current sheet forms as the closed magnetic field is severely 
stretched by the eruption, separating magnetic field lines of opposite 
polarity, and the temporal vacuum around the sheet quickly brings these 
field lines and plasma into the sheet driving reconnection to occur (e.g., 
see {\it Lin et al.} [2003]). In this process it is not necessarily the 
magnetic field of the low Alfv\'{e}n speed that reconnects before that of 
high Alfv\'{e}n speed. But it is true that the Alfv\'{e}n speed inside the 
sheet is low compared to that outside (e.g., see {\it Forbes and Malherbe} 
[1991]).

\section{Summary}
As an efficient energy conversion process in magnetized plasma 
environments, magnetic reconnection takes place in the solar atmosphere, 
magnetosphere, laboratory plasmas, and in other astrophysical contexts. 
Here we have emphasized the solar and magnetospheric context because, so 
far, these are the only two places where the magnetic reconnection 
process on the large scale can be observed or detected directly. We 
compared their morphological features and physical properties, discussed the 
reconnection process as a means of energy conversion and detailed the 
observational consequences. This comparative study reveals several pieces of
important information that help us understand the physics related to the 
large scale energy conversion and release processes that may occur in the 
magnetized plasma environments on other objects in the universe.

The main results we deduced from the comparison performed in the present 
work are summarized as follows:

1. In both solar atmosphere and magnetosphere, reconnection slowly sends 
the magnetic flux and energy from the region of low $V_{A}$ and high 
$\beta$ to that of high $V_{A}$ and low $\beta$, then the magnetic energy 
stored in the latter is quickly released by reconnection causing violent 
eruptions.

2. Energy conversion through magnetic reconnection in both environments 
takes place in current sheets. These sheets form as a result of the 
interaction between the solar wind and the geomagnetic field in the 
magnetosphere, in one case; and in the corona, they develop when the closed 
magnetic field lines are stretched by the eruption.

3. Reconnection takes place in the current sheet when the sheet becomes 
thin enough. In the Earth's magnetotail, measurements by Geotail and 
Cluster indicate that reconnection does not take place until the sheet 
has thinned to the point where not only the ions have become 
``demagnetized", but embedded electron scale effects become important to 
provide the necessary dissipation as well. In the solar CME/flare, it was 
also shown that reconnection commenced when two magnetic fields of 
opposite polarity were pushed against each other, which implies that the 
sheet between the two fields got thinned (e.g., see {\it Lin et al.} 
[2005]). At present, it is not clear how thin the sheet should be, but it 
is usually believed that reconnection begins as a result of the tearing 
mode when the sheet length exceeds $\sim 2\pi$ times its thickness (e.g., 
see {\it Furth et al.} 1963; {\it Priest and Forbes} [2000]).

4. Huge difference between the two contexts in length scale and in 
magnetic field strength yields significant difference between them in 
the absolute values of both amount and rate of energy release during 
eruptions. But if we consider a dimensionless energy release rate, $M_{A}$, 
namely the relative rate of reconnection, then we found that the eruptive 
processes in the different environments typically take place at the same 
rate. This implies that the energy conversion via reconnection on other 
objects in the universe are quite likely to occur at the same $M_{A}$ as 
on the Sun and on the Earth.

5. The most magnificent morphological features observed (detected) in the 
solar eruption (substorms) are the plasma blobs (plasmoids) flowing along 
the respective reconnecting current sheets. The flow of the blob 
observed in several solar eruptions was identified with the reconnection 
outflow, and was used to deduce the Alfv\'{e}n speed in the corona.

6. The blob flows in both directions are well observed, but those 
propagating away from the Sun in solar eruptions and tailward in 
substorms can be recognized more easily than those towards the Sun and 
the Earth. The closed field lines below the current sheet produced by 
magnetic reconnection is the reason responsible for such a discrimination 
of the plasma flow behaviors in both contexts.

7. Plasma blobs or plasmoids could be identified with either the magnetic 
islands caused by the tearing mode instability (turbulence) or the 
reconnection outflow regions surrounded by the slow mode shocks in the 
Petschek-type reconnection process. In both cases, magnetic reconnection 
dissipates the magnetic field in a fairly efficient fashion. 
The high efficiency of the energy conversion results from the 
hyper-resistivity in the former case, and from the slow mode shocks in the 
latter case. More work in theory, observation, and numerical experiments 
are expected to shed light on this issue.


%
%



%
%


%
%

\begin{acknowledgments}
We thank N. Crooker for valuable comments and suggestions. We are also 
grateful for both referee's valuable comments and suggestions that help 
improve this paper. JL thanks ISSI (International Space Science 
Institute, Bern) for the hospitality provided to the members of the team 
on the Role of Current Sheets in Solar Eruptive Events where some of the 
ideas presented in this work have been discussed. JL's work at YNAO was 
supported by the Ministry of Science and Technology of China under the 
Program 973 grant 2006CB806303, by the National Natural Science 
Foundation of China under the grant 40636031, and by the Chinese Academy 
of Sciences under the grant KJCX2-YW-T04 to YNAO, and he was supported by 
NASA grant NNX07AL72G when visiting CfA. SRC acknowledges support from 
NASA grant NNG04GE77G. CJF acknowledges support from NASA grants NNX08AD11G,
NNG05GC75G, and NNG06GDD41G. {\it SOHO} is a joint mission of the European 
Space Agency and the US National Aeronautics and Space Administration.
\end{acknowledgments}

%
%
%
%
%
%
%
%


\end{article}

%
%
%

------------------------------------------------------------------------ 
%
%
%
\begin{table}
\caption{Some important parameters for plasmas in various circumstances}
\begin{flushleft}
\begin{tabular}{ccccccccc}
\tableline
& \multicolumn{2}{c}{$B$\tablenotemark{a}} & \multicolumn{2}{c}{$n_{e}$ 
(cm$^{-3}$)} 
& \multicolumn{2}{c}{$V_{A}$ (10$^{3}$ km s$^{-1}$)}
& \multicolumn{2}{c}{$\beta$\tablenotemark{b}} 
\\
\tableline
Sun\tablenotemark{c} & AR & Ph & AR & Ph & AR & Ph & AR & Ph\\
& 10$^{2}$ & 10$^{3}$ & 10$^{10}$ & 10$^{17}$ & 1 & $<0.01$ & $3.5\times 
10^{-3}$ & $>2$\\
\tableline
Earth & Dayside\tablenotemark{d} & Tail lobe\tablenotemark{e} & Dayside & 
Tail lobe & Dayside & Tail lobe & Dayside & Tail lobe\\
& 100 & 20 & 100 & 0.01 & 0.22  & 4.4 & 0.34 & 
$8.6\times 10^{-3}$\\
 & NECPS\tablenotemark{f} & MDPS\tablenotemark{g} & NECPS & 
MDPS & NECPS & MDPS & NECPS & MDPS\\
& 5 & 5 & 0.35 & 0.25 & 0.78  & 0.16 & 25 & 1\\
\tableline
\end{tabular}
\end{flushleft}
\tablenotetext{a}{Units of magnetic field strength in the corona is G, and 
that in the magnetosphere is nT.}
\tablenotetext{b}{Plasma $\beta$, the ratio of the gas pressure to the 
magnetic pressure, $8\pi n_{e}kT/B^{2}$. Here temperature $T=10^{6}$ K in 
the corona and in the magnetopause, and $k$ is the Boltzmann 
constant.}
\tablenotetext{c}{Parameters for the solar atmosphere. AR indicates the 
coronal base over the active region, and Ph stands for the photosphere.}
\tablenotetext{d}{Parameters are measured at about 10
Earth radii close to the noon direction. (e.g., see also {\it Paschmann et 
al.} [1986]).}
\tablenotetext{e}{For comparison with solar flares, near tail (around 20 
$R_{E}$ or so) data are used here: $B=20$ nT, $T=10^{7}$, and 
$n_{e}=0.01$ cm$^{-3}$.}
\tablenotetext{f}{Information from statistical survey by AMPTE/IRM in 
1986 of ion/proton moments in the near-Earth (between -9 
and -18 $R_{E}$ of GSM distance) central plasma sheet (NECPS) [{\it  
Baumjohann et al.} 1989]. Note: What {\it Baumjohann et al.} [1989] used 
was the ion density. So by quoting this as $n_{e}$ we are assuming an 
electron-proton population and charge neutrality. A typical ion 
temperature in NECPS is $T_{i}=5\times 10^{7}$ K, and the electron 
temperature is around $T_{e}=10^{7}$ K.} 
\tablenotetext{g}{Information for the middle distance plasma sheet (MDPS) 
from {\it  Slavin et al.} [1985].}
\label{tbl:VA} 
\end{table}

\end{document}